\documentclass[a4paper,fleqn,usenatbib]{mnras}

\usepackage[T1]{fontenc}
\usepackage{ae,aecompl}

\usepackage{graphicx}	
\usepackage{amsmath,amsfonts,amssymb}
\usepackage{makeidx}
\usepackage{epstopdf} 
\usepackage{booktabs,caption,fixltx2e}
\usepackage[flushleft]{threeparttable}

\usepackage{hyperref}
\usepackage{multirow}
\usepackage{rotating}
\usepackage{color}
\usepackage{soul}
\usepackage{ulem}

\def\arcmin{\hbox{$^\prime$}}
\def\arcsec{\hbox{$^{\prime\prime}$}}

\newcommand{\spitzer}{{\it Spitzer}}

\newcommand{\LCDM}{\mbox{$\Lambda$CDM}}

\newcommand{\ltsima}{$\; \buildrel < \over \sim \;$}
\newcommand{\ltsim}{\lower.5ex\hbox{\ltsima}}

\newcommand{\be}{\begin{equation}}
\newcommand{\ee}{\end{equation}}
\newcommand{\bea}{\begin{eqnarray}}
\newcommand{\eea}{\end{eqnarray}}

\newcommand{\degs}{deg$^2$}

\newcommand{\project}[1]{{\sffamily #1}}
\newcommand{\thisplain}{emcee}
\newcommand{\emcee}{\project{\thisplain}}

\def\mstar{$m^*$}

\def\hennigcgall{$3.59^{+0.20}_{-0.18}$}
\def\hennigcgred{$5.37^{+0.27}_{-0.24}$}

\def\CGallevosol  {$c_{\rm g,all}=5.01^{+1.02}_{-1.29} \times (1+z)^{-1.21 \pm 0.59}$}
\def\CGallevosolslope  {$-1.21 \pm 0.59$}
\def\CGallevosignificance  {$2.05$}

\def\CGredevosol  {$c_{\rm g,\rm red}=6.03^{+1.23}_{-1.55} \times (1+z)^{-1.74^{+0.62}_{-0.64}}$}
\def\CGredevosolslope  {$-1.74^{+0.62}_{-0.64}$}
\def\CGredevosignificance  {$2.81$}

\def\RPallstacksol{$2.36^{+0.38}_{-0.35}$} 
\def\LFCGCORRECTIONALL{$c_{\rm corr}   =  2.36$}

\def\RPredstacksol{$2.84^{+0.40}_{-0.37}$}
\def\LFCGCORRECTIONRED{$c_{\rm corr}   =  2.84$}

\def\phiallsol   {$\phi^*_{\rm all}    =2.24^{+0.23}_{-0.20}$}

\def\alphaallsol {$\alpha_{\rm all}    =-1.06^{+0.04}_{-0.03}$}
\def\chiallsol   {$\chi^2_{\rm all,red}=2.96$}

\def\phiredsol   {$\phi^*_{\rm rs}    =2.21^{+0.16}_{-0.15}$}

\def\alpharedsol {$\alpha_{\rm rs}    =-0.80^{+0.04}_{-0.03}$}
\def\chiredsol   {$\chi^2_{\rm rs,red}=1.31$}

\def\mstarevointer{$0.11^{+0.12}_{-0.13}$}
\def\mstarevoslope{$-0.12^{+0.30}_{-0.29}$}

\def\phievoslopemfix{$-0.81^{+0.34}_{-0.34}$}
\def\phiallsignificancemfix{$2.38$}

\def\alphaevointer{$-1.05^{+0.05}_{-0.05}$}
\def\alphaevoslope{$-0.04^{+0.14}_{-0.14}$}
\def\alphaallsignificance{$0.29$}

\def\mstarevointerallfree{$0.25^{+0.41}_{-0.41}$}
\def\mstarevoslopeallfree{$0.33^{+1.21}_{-1.20}$}

\def\mstarevointerallfreered{$0.54^{+0.17}_{-0.17}$}
\def\mstarevoslopeallfreered{$0.20^{+0.37}_{-0.37}$}

\def\HONNORMPIVOT{$223.87^{+5.22}_{-10.06}$}

\def\alphaslopebootstrap{$0.15$}
\def\alphaslopeerrbootstrap{$0.14$}
\def\alphaslopesignificancebootstrap{$1.08$}

\def\HONevointer{$0.06^{+0.06}_{-0.05}$}
\def\HONevoslope{$-0.80^{+0.38}_{-0.38}$}
\def\HONallEvolSignificance{$2.11$}

\def\mstarredevoslope{$-0.25^{+0.17}_{-0.18}$}

\def\phiredevoslopemfix{$-0.47^{+0.30}_{-0.29}$}
\def\phiredsignificancemfix{$1.57$}

\def\alpharedevointer{$-0.87^{+0.04}_{-0.04}$}
\def\alpharedevoslope{$0.21^{+0.09}_{-0.10}$}
\def\alpharedsignificance{$2.10$}

\def\onessims{$3.09 \pm 0.09$}

\newcommand{\CTIO}{$^{1}$}
\newcommand{\Munich}{$^{2}$}
\newcommand{\ExcellenceCluster}{$^{3}$}
\newcommand{\MPE}{$^{4}$}
\newcommand{\CfA}{$^{5}$}
\newcommand{\UHawaii}{$^{6}$}
\newcommand{\colby}{$^{7}$}
\newcommand{\Harvard}{$^{8}$}
\newcommand{\UFlorida}{$^{9}$}
\newcommand{\UMontreal}{$^{10}$}
\newcommand{\MIT}{$^{11}$}
\newcommand{\Melbourne}{$^{12}$}
\newcommand{\STSI}{$^{13}$}

\title{Galaxy Populations in the 26 most massive Galaxy Clusters in the South Pole Telescope SZE Survey}
\author[A.~Zenteno et al.]{A.~Zenteno \CTIO$^,$\Munich, 
J. J.~Mohr\Munich$^,$\ExcellenceCluster$^,$\MPE,
S.~Desai\Munich$^,$\ExcellenceCluster,
B.~Stalder\CfA$^,$\UHawaii,
A.~Saro\Munich$^,$\ExcellenceCluster,
J. P.~Dietrich\Munich$^,$\ExcellenceCluster,
\newauthor
M.~Bayliss\colby$^,$\Harvard,
S.~Bocquet\Munich$^,$\ExcellenceCluster,
I.~Chiu\Munich$^,$\ExcellenceCluster,
A. H.~Gonzalez\UFlorida,
C.~Gangkofner\Munich$^,$\ExcellenceCluster,
\newauthor
N.~Gupta\Munich$^,$\ExcellenceCluster,
J.~Hlavacek-Larrondo\UMontreal,
M.~McDonald\MIT,
C. Reichardt\Melbourne, and
A. Rest\STSI
\\
\CTIO Cerro Tololo Inter-American Observatory, Casilla 603, La Serena, Chile \\
\Munich Faculty of Physics, Ludwig-Maximilians-Universit\"at, Scheinerstr.\ 1, 81679 Munich, Germany \\
\ExcellenceCluster Excellence Cluster Universe, Boltzmannstr.\ 2, 85748 Garching, Germany \\
\MPE Max-Planck-Institute for Extraterrestril Physics, Giessenbachstr.\ 85748 Garching, Germany \\
\CfA Harvard-Smithsonian Center for Astrophysics, 60 Garden Street, Cambridge, MA 02138 \\
\UHawaii Institute for Astronomy, University of Hawaii at Manoa, Honolulu, HI 96822, USA \\
\colby Department of Physics \& Astronomy, Colby College, 5800 Mayflower Hill, Waterville, Maine 04901 \\
\Harvard Department of Physics, Harvard University, 17 Oxford Street, Cambridge, MA 02138 \\
\UFlorida Department of Astronomy, University of Florida, Gainesville, FL 32611 \\
\UMontreal D\'epartement de Physique, Universit\'e de Montr\'eal, C.P. 6128, Succ. Centre-Ville, Montr\'eal, Qu\'ebec H3C 3J7, Canada \\
\MIT Kavli Institute for Astrophysics and Space Research, Massachusetts Institute of Technology, 77 Massachusetts Avenue, Cambridge, MA 02139 \\
\Melbourne School of Physics, University of Melbourne, Parkville, VIC 3010, Australia \\
\STSI Space Telescope Science Institute, 3700 San Martin Dr., Baltimore, MD 21218 
}

\date{Accepted XXX. Received YYY; in original form ZZZ}

\pubyear{2016}

\begin{document}
\label{firstpage}
\pagerange{\pageref{firstpage}--\pageref{lastpage}}
\maketitle
%
\begin{abstract}

We present  a study of the  optical properties of the  26 most massive
galaxy clusters selected within  the SPT-SZ 2500~deg$^2$ survey.  This
Sunyaev-Zel'dovich effect  selected sample  spans a redshift  range of
0.10 $<  z <$ 1.13.   We measure  the galaxy radial  distribution, the
luminosity function (LF),  and the halo occupation  number (HON) using
optical data with a typical depth of $m^*+2$ within the band that lies
just  redward of  the 4000~\AA\  break at  the cluster  redshift.  The
stacked  radial profiles  are  consistent  with a  Navarro-Frenk-White
profile with  a concentration  of \RPredstacksol~for the  red sequence
and  \RPallstacksol~for the  total population.   Stacking the  data in
multiple  redshift bins  shows a  hint  of redshift  evolution in  the
concentration when  both the total  population is used, and  when only
red sequence  galaxies are used (at  \CGallevosignificance$\sigma$ and
\CGredevosignificance$\sigma$, respectively).  The  stacked LF shows a
faint end slope \alphaallsol\ for  the total and \alpharedsol\ for the
red sequence population.  The redshift evolution of the characteristic
magnitude \mstar\ is found to  be consistent with a passively evolving
Composite Stellar Population (CSP) model over the full redshift range.
By adopting the CSP model predictions for the characteristic magnitude
\mstar,  we explore  the redshift  evolution  in the  faint end  slope
$\alpha$  and   characteristic  galaxy  density  $\phi^*$.    We  find
$\alpha$\ for the total population  to be consistent with no evolution
(\alphaallsignificance$\sigma$), while  evidence of evolution  for the
red          galaxies          is          mildly          significant
(\alphaslopesignificancebootstrap$-$\alpharedsignificance$\sigma$),
with a  steeper faint end  at low redshifts.   The data show  that the
density  $\phi^*/E^2(z)$  of  galaxies with  characteristic  magnitude
$m^*$  decreases  with  redshift,  in tension  with  the  self-similar
expectation at a \phiallsignificancemfix  $\sigma$ level for the total
population,  when  \mstar\  is  fixed  to  the  model.   The  measured
HON--mass  relation for  our sample-wide  redshift range  has a  lower
normalization  than previous  studies at  low redshift.   Finally, our
data  support  HON  redshift evolution  at  a  \HONallEvolSignificance
$\sigma$  level, with  clusters  at higher  redshift containing  fewer
galaxies per unit mass to $m^*+3$ than their low-z counterparts.

\end{abstract}

\begin{keywords}
galaxies:  clusters:  general   --  galaxies:  evolution  --
  galaxies: formation -- cosmology: observations -- Sunyaev-Zel'dovich
  Effect 
\end{keywords}



\section{Introduction}
\label{sec:intro}

Clusters have long  been recognised as important  laboratories for the
study       of       galaxy        formation       and       evolution
\citep[e.g.,][]{spitzer1951,dressler80,butcher84,depropris03,andreon10}.
With the  advent of  the new generation  of mm-wave  survey telescopes
like  the  South  Pole   Telescope  \citep{carlstrom11},  the  Atacama
Cosmology     Telescope     \citep[ACT,][]{fowler07}    and     Planck
\citep{planck11-13}, it has become  possible to select galaxy clusters
over  large  fractions of  the  extragalactic  sky using  the  thermal
Sunyaev-Zel'dovich effect (SZE), which arises from the inverse Compton
scattering of  CMB photons off  the hot electrons in  the intracluster
medium \citep{sunyaev72}.  For the SPT-SZ arcminute angular resolution
2500~deg$^2$ survey, it has been demonstrated that the cluster samples
selected   using   this   signature   are  close   to   mass   limited
\citep{reichardt13},   extend   to    at   least   redshift   $z=1.47$
\citep{bayliss14}  and  have  purity   exceeding  95\%  from  the  SZE
selection  alone  \citep{song12b,bleem15b}.   These  cluster  samples,
selected using  cluster gas  signatures as  opposed to  cluster galaxy
signatures, are ideal  for evolutionary studies of  the cluster galaxy
populations.

By studying  the evolution of  the cluster galaxy  luminosity function
(LF)  we can  address  the changes  in the  cluster  populations in  a
statistical  manner.   It  has  been  shown,  that  while  the  bright
population  is consistent  with  a passive  evolution  of the  stellar
population,  the  faint-end  of  the red  sequence  LF  (rLF)  becomes
increasingly         shallow        at         higher        redshifts
\citep[e.g][]{delucia07b,gilbank08,rudnick09}.  Furthermore,  the same
studies hint  at a weak  correlation of the luminosity  function faint
end slope $\alpha$ with mass.  At the same time, previous studies have
shown that the Halo Occupation Number (HON), or the integral of the LF
per    unit   mass,    seems   to    be   invariant    with   redshift
\citep{lin04a,lin06}, which points to continuous galaxy transformation
within  the cluster.   This transformation  can also  be tracked  as a
function  of the  radius,  using the  concentration  evolution of  the
different species.   Literature values at different  redshifts seem to
indicate no evolution  when all galaxies within the  virial radius are
considered   \citep[e.g.,][]{carlberg97,capozzi12},   and  while   the
expectation  is  that  the  brightest  red  sequence  galaxies,  which
dominate the bright-end of the LF, would be more concentrated than the
fainter  component, it  is not  known whether  this effect  is present
already at high  redshift.  All these components are also  used in the
framework      of      the      Halo      Occupation      Distribution
\citep[HOD;][]{berlind03},  which describes  how  galaxies occupy  the
cluster  as a  function of  the location,  velocity distribution,  and
luminosity.

In this work, we extract  the radial distribution, luminosity function
and the  HON of  galaxies in  SZE selected  cluster sample  to address
cluster galaxy evolution questions cleanly within a uniformly selected
sample of  the most massive  clusters in  the Universe.  This  work is
complementary  to  that  of   \citet{hennig16},  a  study  of  optical
properties of a larger sample of 74 SZE selected clusters with a lower
average mass within a limited sky area of 200 deg$^2$.  Our goal is to
study how the galaxy components,  separated into the red subsample and
the full sample within the virial radius, change over cosmic time.  By
making reference  to previous  studies that have  been carried  out on
X-ray and optically selected cluster  samples, we have the opportunity
to  begin to  address  the  importance of  sample  selection in  these
studies.

The paper  is organized  as follows:  Section~\ref{sec:data} describes
the observations  and data  reduction. In  Section~\ref{sec:tools}, we
describe  our  tools  and  the  simulations used  to  test  them.   In
Section~\ref{sec:results} we present the main  results of the study of
the galaxy  populations in  the SPT  selected massive  cluster sample.
Conclusions      of     this      study      are     presented      in
Section~\ref{sec:conclusions}.   Magnitudes  are   quoted  in  the  AB
system.    We  assume   a   flat,  \LCDM\   cosmology   with  $H_0   =
70.2$~km~s$^{-1}$~Mpc$^{-1}$,  and matter  density $\Omega_{\rm  M}$ =
0.272, according  to WMAP7  + BAO  + H0 data  (Komatsu et  al.  2011).
Masses   are   defined   as    $M_{\Delta,\rm   crit}=   \frac{4   \pi
  r_{\Delta}^3}{3}    \Delta\rho_{\rm    crit}$,   where    $\rho_{\rm
  crit}=3H^2/8\pi G$ is the critical density of the Universe.

\section{Observations and Data Reduction}
\label{sec:data}
In this work  we use a sample  of the most massive  galaxy clusters in
the total 2500  deg$^2$ SPT survey area that  was originally presented
in \citet{williamson11}.   The sample  consists of 26  galaxy clusters
with      masses     $M_{\rm      200,crit}      >     1.2      \times
10^{15}h_{70}^{-1}$M$_\odot$ extending to redshift $z=1.13$.

The optical photometric and spectroscopic data used in this paper come
from multiple observatories and they have been processed using several
pipelines.   The data  reductions for  a  portion of  the dataset  are
outlined  in several  papers \citep{high10,williamson11,song12b}.   In
the following subsections we summarize the data and the processing and
calibration.

\begin{figure}
\begin{center}
\includegraphics[width=0.49\textwidth]{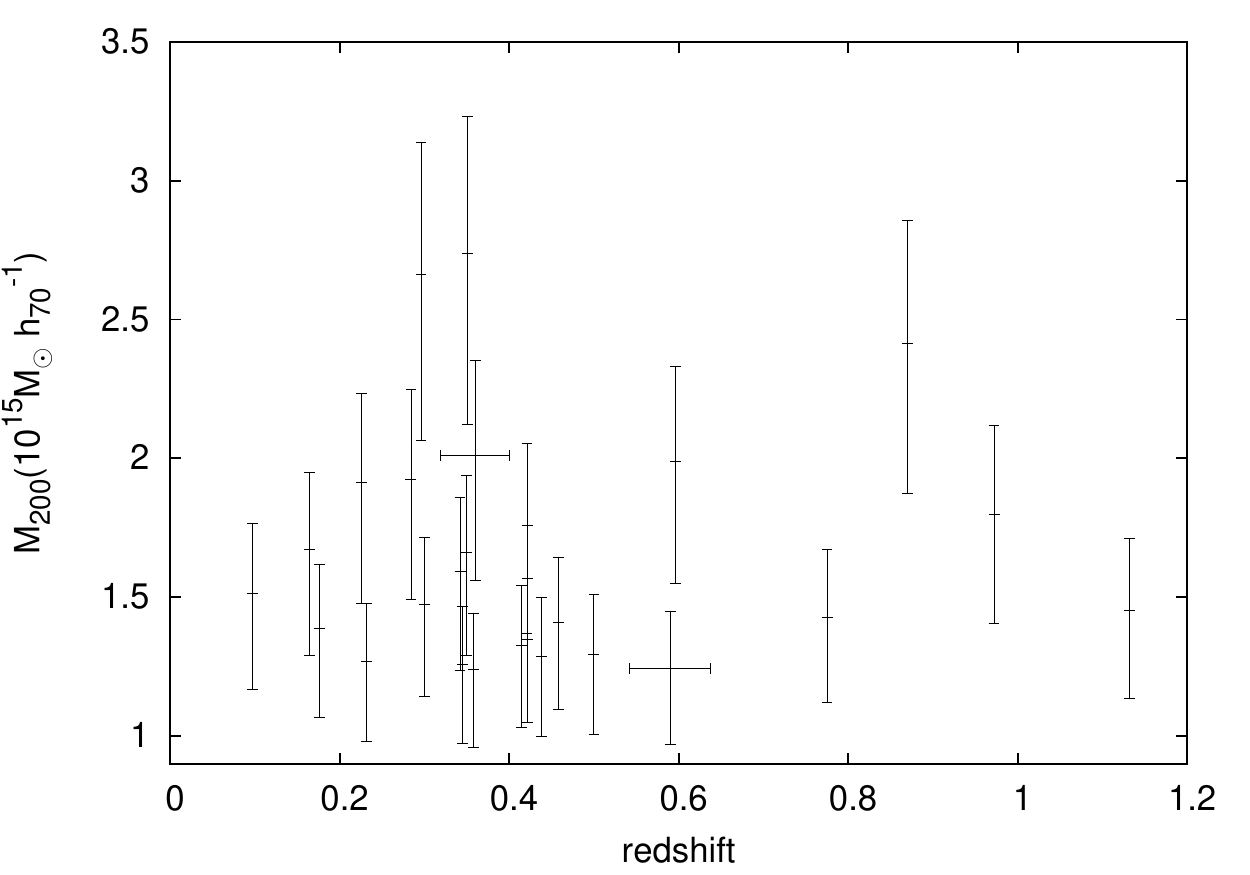}
\caption[Mass-redshift distribution of \protect\citet{williamson11} sample.]{Mass-redshift distribution of \protect\citet{williamson11} sample.  Masses where obtained from \protect\citet{bocquet15} and redshifts from \protect\citet{bleem15b}.}
\end{center}
\end{figure}

\subsection{mm-wave Observations}
\label{sec:sz}
The clusters presented here are the most massive systems in the SPT-SZ
survey  area,  which consists  of  a  contiguous 2500  \degs\,  region
defined  by the  boundaries  $20^{\mathrm{h}}  \le \mathrm{R.A.}   \le
24^{\mathrm{h}}, 0^{\mathrm{h}} \le \mathrm{R.A.}  \le 7^{\mathrm{h}}$
and $-65^\circ  \le {\rm decl.}  \le -40^\circ$.  Mass  estimation for
the clusters  has been  carried out  in a  staged manner,  first using
simulations \citep{vanderlinde10},  and then  using a small  number of
X-ray $Y_X$ measurements  \citep{benson13,reichardt13}. For details on
the SPT  data processing  there are several  papers that  describe the
method in detail \citep{staniszewski09,vanderlinde10,shirokoff11}.

\subsection{Redshifts and Cluster Masses}
\label{sec:redshifts}
Cluster   redshifts   first   appeared    in   the   discovery   paper
\citep{williamson11},   but   since  then   additional   spectroscopic
redshifts   have   become  available   for   six   of  these   systems
\citep{song12b,ruel14,sifon13,planck11-9}.   Where   possible  we  use
spectroscopic    redshifts.    The    redshifts    are    listed    in
Table~\ref{tab:3}.   This table  contains the  SPT cluster  name (with
reference to  other names where they  exist), the SPT sky  position of
the  cluster (R.A.  and  Decl.), the  redshift  (with two  significant
digits if a  photo-z and with three if a  spectroscopic redshift), the
SPT S/N, $\xi$,  the estimated cluster mass, the virial  radius in arc
minutes, and  the Brightest  Cluster Galaxy  (BCG) position  (R.A. and
Decl.).

Although   \citet{williamson11}   reported  $M_{200,\rm   mean}$   and
$M_{500,\rm crit}$  masses for each  cluster, we update the  values of
$M_{500,\rm crit}$  using the \citet{bocquet15} code.   We convert the
$M_{500,\rm crit}$ to $M_{200,\rm  crit}$ (hereafter $M_{200}$), using
a   Navarro-Frenk-White   profile  \citep[NFW;][]{navarro97}   and   a
concentration-mass relation from \citet{duffy08}.  Masses are shown in
Table~\ref{tab:3} along with the corresponding angular projected radii
at  which  the  cluster density  reaches  200$\times\rho_{\rm  crit}$,
hereafter $r_{200}$, given the assumed cosmology.

\begin{center}
\begin{table*}
    \caption{SPT Cluster List: $^1$ACT-CL J0102-4915; $^2$RXC J0232.2-4420;$^3$ABELL S0295, ACT-CL J0245-5302;$^4$ACT-CL J0438-5419;$^5$ABELL 3396, RXC J0628.8-4143;$^6$ABELL S0592, RXC J0638.7-5358. ACT-CL J0638-5358;$^7$ABELL 3404, RXC J0645.4-5413, ACT-CL J0645-5413;$^8$Bullet, RXC J0658.5-5556, ACT-CL J0658-5557;$^9$RXC J2023.4-5535;$^{10}$RXC J2031.8-4037;$^{11}$ABELL 3827, RXC J2201.9-5956;$^{12}$ABELL S1063, RXC J2248.7-4431;$^{13}$ABELL S1121; $^{\rm p}$Photometric redshift are accurate to $\sigma_z/(1+z) \approx 2\%-3\%$  }\label{tab:3}
    \begin{tabular}{lrrlrrrrr}
    \hline \hline   
\multicolumn{1}{c}{Object Name}& 
\multicolumn{1}{c}{R.A.} & 
\multicolumn{1}{c}{Decl.} & 
\multicolumn{1}{c}{$z$} & 
\multicolumn{1}{c}{S/N} & 
\multicolumn{1}{c}{$M_{200}$} & 
\multicolumn{1}{c}{$R_{200}$} & 
\multicolumn{1}{c}{$\textrm{R.A.}_{\rm BCG}$} &
\multicolumn{1}{c}{$\textrm{Decl.}_{\rm BCG}$}
\\ 
\multicolumn{1}{c}{} &
\multicolumn{1}{c}{[deg]} &
\multicolumn{1}{c}{[deg]} &
\multicolumn{1}{c}{} &
\multicolumn{1}{c}{[$\xi$]} &
\multicolumn{1}{c}{[$10^{14}h_{70}^{-1}$M$_\odot$]} &
\multicolumn{1}{c}{[$\arcmin$]} &
\multicolumn{1}{c}{[deg]} & 
\multicolumn{1}{c}{[deg]} \\ \hline \hline
SPT-CL J0040-4408 & 10.202 & $-44.131$ & 0.350 & 19.34 & $16.61^{+2.78}_{-3.73}$ & 7.33 & 10.2083 &  $-44.1305$ \\ 
SPT-CL J0102-4915$^1$ & 15.728 & $-49.257$ & 0.870 & 39.91 & $24.14^{+4.44}_{-5.40}$ & 4.34 & 15.7221 &  $-49.2530$ \\ 
SPT-CL J0232-4421$^2$ & 38.070 & $-44.351$ & 0.284 & 23.96 & $19.26^{+3.24}_{-4.34}$ & 9.09 & 38.0680 &  $-44.3466$ \\ 
SPT-CL J0234-5831 & 38.670 & $-58.520$ & 0.415 & 14.66 & $13.25^{+2.17}_{-2.94}$ & 5.96 & 38.6762 &  $-58.5235$ \\ 
SPT-CL J0243-4833 & 40.910 & $-48.557$ & 0.500 & 13.90 & $12.96^{+2.14}_{-2.88}$ & 5.15 & 40.9120 &  $-48.5607$ \\ 
SPT-CL J0245-5302$^3$ & 41.378 & $-53.036$ & 0.300 & 15.95 & $14.75^{+2.41}_{-3.31}$ & 7.96 & 41.3535 &  $-53.0292$ \\ 
SPT-CL J0254-5856 & 43.563 & $-58.949$ & 0.438 & 14.13 & $12.87^{+2.11}_{-2.87}$ & 5.67 & 43.5365 &  $-58.9717$ \\ 
SPT-CL J0304-4401 & 46.064 & $-44.030$ & 0.458 & 15.69 & $14.10^{+2.35}_{-3.14}$ & 5.65 & 46.0878 &  $-44.0438$ \\ 
SPT-CL J0411-4819 & 62.811 & $-48.321$ & 0.422 & 15.26 & $13.47^{+2.20}_{-2.97}$ & 5.92 & 62.8154 &  $-48.3174$ \\ 
SPT-CL J0417-4748 & 64.340 & $-47.812$ & 0.59$^{\rm p}$ & 14.24 & $12.43^{+2.06}_{-2.73}$ & 4.52 & 64.3463 &  $-47.8132$ \\ 
SPT-CL J0438-5419$^4$ & 69.569 & $-54.321$ & 0.421 & 22.88 & $17.59^{+2.94}_{-3.90}$ & 6.48 & 69.5738 &  $-54.3223$ \\ 
SPT-CL J0549-6205 & 87.326 & $-62.083$ & 0.36$^{\rm p}$ & 25.81 & $20.12^{+3.40}_{-4.50}$ & 7.64 & 87.3332 &  $-62.0870$ \\ 
SPT-CL J0555-6406 & 88.851 & $-64.099$ & 0.345 & 12.72 & $12.59^{+2.07}_{-2.85}$ & 6.76 & 88.8537 &  $-64.1055$ \\ 
SPT-CL J0615-5746 & 93.957 & $-57.778$ & 0.972 & 26.42 & $17.96^{+3.21}_{-3.92}$ & 3.66 & 93.9656 &  $-57.7801$ \\ 
SPT-CL J0628-4143$^5$ & 97.201 & $-41.720$ & 0.176 & 13.89 & $13.87^{+2.30}_{-3.19}$ & 12.17 & 97.2073 &  $-41.7269$ \\ 
SPT-CL J0638-5358$^6$ & 99.693 & $-53.974$ & 0.226 & 22.69 & $19.14^{+3.21}_{-4.35}$ & 10.95 & 99.6882 &  $-53.9730$ \\ 
SPT-CL J0645-5413$^7$ & 101.360 & $-54.224$ & 0.164 & 18.32 & $16.73^{+2.78}_{-3.81}$ & 13.78 & 101.3725 &  $-54.2267$ \\ 
SPT-CL J0658-5556$^8$ & 104.625 & $-55.949$ & 0.296 & 39.05 & $26.64^{+4.75}_{-5.99}$ & 9.79 & 104.6777 &  $-55.9765$ \\ 
SPT-CL J2023-5535$^9$ & 305.833 & $-55.590$ & 0.232 & 13.63 & $12.70^{+2.08}_{-2.88}$ & 9.34 & 305.9069 &  $-55.5696$ \\ 
SPT-CL J2031-4037$^{10}$ & 307.960 & $-40.619$ & 0.342 & 17.52 & $15.95^{+2.65}_{-3.57}$ & 7.36 & 307.9492 &  $-40.6151$ \\ 
SPT-CL J2106-5844 & 316.515 & $-58.744$ & 1.132 & 22.22 & $14.51^{+2.59}_{-3.14}$ & 3.11 & 316.5190 &  $-58.7412$ \\ 
SPT-CL J2201-5956$^{11}$ & 330.462 & $-59.944$ & 0.097 & 15.26 & $15.13^{+2.52}_{-3.47}$ & 21.36 & 330.4723 &  $-59.9453$ \\ 
SPT-CL J2248-4431$^{12}$ & 342.181 & $-44.527$ & 0.351 & 42.36 & $27.37^{+4.94}_{-6.13}$ & 8.64 & 342.1832 &  $-44.5307$ \\ 
SPT-CL J2325-4111$^{13}$ & 351.294 & $-41.194$ & 0.358 & 12.50 & $12.39^{+2.04}_{-2.81}$ & 6.53 & 351.2988 &  $-41.2034$ \\ 
SPT-CL J2337-5942 & 354.347 & $-59.703$ & 0.775 & 20.35 & $14.28^{+2.43}_{-3.08}$ & 3.93 & 354.3652 &  $-59.7013$ \\ 
SPT-CL J2344-4243 & 356.176 & $-42.719$ & 0.596 & 27.44 & $19.88^{+3.42}_{-4.38}$ & 5.24 & 356.1830 &  $-42.7200$ \\ 
 \hline
 \hline
    \end{tabular}
    \end{table*}  
\end{center}

\subsection{Optical Imaging}
\label{sec:imaging}
The present  cluster sample has  been imaged with  several instruments
and  telescopes,  and  with  different goals  in  mind:  from  shallow
photometry for  photometric redshift estimations to  deep observations
for weak lensing analysis (see  Table~\ref{tab:ocameras} for a list of
the  telescopes/instruments  used).   Those   observations  led  to  a
heterogeneous dataset.   To `homogenize'  the sample  we set  a common
luminosity limit of $m^*+2$ (\mstar~being the characteristic magnitude
of  the   luminosity  function)   at  10$\sigma$  for   each  cluster,
re-observing several of them in order  to achieve this goal.  The data
reduction is performed  using three different pipelines,  and they are
summarized below.

\subsubsection{Mosaic2 Imager}
The Mosaic2 imager was a prime focus camera on the Blanco 4m telescope
until 2012 when it was decommissioned  in favour of the new wide field
DECam imager.  Mosaic2 contained eight 2048$\times$4096 CCD detectors.
However, one  of the amplifiers of  CCD \# 4 had  been non-operational
for the  last three years  coinciding with these  observations.  Given
the fast optics  at the prime focus on the  Blanco, the pixels subtend
0.27\arcsec\ on the sky. Total field of view is 36.8\arcmin\ on a side
for a total solid angle per exposure of $\sim$0.4 \degs.  More details
on   the  Mosaic2   imager   can   be  found   in   the  online   CTIO
documentation\footnote{
  http://www.ctio.noao.edu/mosaic/manual/index.html}.

The data from the Mosaic2 imager  for this analysis is reduced using a
development version of the Dark Energy Survey Data Management Pipeline
(DESDM)~\citep{desai12}.   In the  DESDM pipeline  the data  from each
night   first  undergoes   detrending   corrections,  which   includes
cross-talk  correction,   overscan  correction,  trimming,   and  bias
subtraction,  as well  as fringe  corrections for  $i$ and  $z$ bands.
Astrometric calibration is done using {\tt SCAMP}~\citep{bertin06} and
using the USNO-B catalog as the astrometric reference.  Co-addition is
done  using  {\tt  SWARP}~\citep{bertin02}. The  single  epoch  images
contributing  to the  coadd are  brought to  a common  zeropoint using
stellar  sources common  to pairs  of images.   The final  photometric
calibration  of the  coadd images  is  carried out  using the  stellar
color-color  locus,   with  reference  to  the   median  SDSS  stellar
locus~\citep{covey07},  as  a  constraint  on  the  zeropoint  offsets
between neighboring  bands, while the absolute  calibration comes from
2MASS~\citep{skrutskie06}.

Mosaic2 data has  been acquired over the period of  2005 to 2012, both
for the Blanco Cosmology Survey \citep[BCS\footnote{The BCS was a NOAO
    Large Survey project that covered  $\sim$80 deg$^2$ over 60 nights
    between  2005  and 2008;}][]{desai12}  and  for  the SPT  targeted
cluster followup.   A detailed  description of the  image corrections,
calibration and typical photometric and astrometric quality appears in
\citet{desai12}.

\subsubsection{WFI, IMACS, and Megacam}
Clusters  outside  the  BCS  footprint  were  observed  using  various
instruments, including WFI, IMACS and Megacam.  For such observations,
the strategy adopted was to adjust  the exposure time to reach a depth
of  $0.4L^*  (m^*+1)$  at  $8\sigma$, to  obtain  robust  red-sequence
photometric redshifts \citep{bleem15b}.   This study required somewhat
deeper imaging than this  photometric redshift estimation strategy, so
the Wide Field Imager (WFI) on the MPG 2.2-meter telescope at La Silla
was  used to  acquire deeper  imaging in  $B-$, $V-$,  $R-$, and  $I-$
filters.     The   initial    imaging   from    IMACS   on    Magellan
\citep{dressler03,osip08}  was typically  deep enough  to use  in this
study, and did not require  additional observations.  We also use $g$,
$r$, and  $i$ band data acquired  with the Megacam imager  on Magellan
\citep{mcleod98}  for   an  ongoing   cluster  weak   lensing  program
\citep[][Dietrich et al. in prep]{high12}.

The processing of  the WFI and IMACS data were  done with the PHOTPIPE
pipeline \citep{rest05a,garg07,miknaitis07}.  WFI data were calibrated
in a procedure analogous to the  Mosaic2 data.  The colors of stars in
the  science data  were calibrated  via the  Stellar Locus  Regression
\citep[SLR; e.g.,  ][]{high09} technique  to a stellar  sequence locus
generated from a catalog of synthetic stellar spectra from the PHOENIX
library \citep{brott05}.   The synthetic stellar locus  was calculated
in the WFI  instrument magnitude system using  CCD, filter, telescope,
and atmospheric throughput measurements.  As  with the other data, the
absolute  calibrations  were  measured  with respect  to  2MASS  point
sources in each field.

The Megacam data reduction was carried out at the Smithsonian
Astrophysical Observatory  (SAO) Telescope  Data Center using  the SAO
Megacam  reduction  pipeline,  and   also  calibrated  using  the  SLR
technique.  See \citet{high12} for  a more detailed description of the
observation strategy and data processing.

\subsubsection{FORS2}
For two  clusters at $z  = 0.87$  and $z =  1.132$ in this  sample, we
acquired VLT/FORS2 data in $b-$, $I-$, and $z-$band under program Nos.
087.A-0843 and  088.A-0796(A) (PI Bazin), 088.A-0889(A,B,C)  (PI Mohr)
and 286.A-5021(A) (DDT, PI  Carlstrom).  Observations were carried out
in queue mode, and were  in clear, although generally not photometric,
conditions.  The  nominal exposure times  for the different  bands are
480\,s ($b$),  2100\,s ($I$), 3600\,s  ($z$).  These were  achieved by
coadding  dithered  exposures with  160\,s  ($b$),  175\,s ($I$),  and
120\,s ($z$). Deviations  from the nominal exposure  times are present
for some fields due to  repeated observations when conditions violated
specified  constraints  or  when  observing  sequences  could  not  be
completed during  the semester  for which  they were  allocated.  Data
reduction  and  calibration  was  performed with  the  THELI  pipeline
\citep{erben05,schirmer13}.     Twilight   flats    were   used    for
flatfielding. The $I-$  and $z-$band data were  defringed using fringe
maps made with  night sky flats constructed from  the data themselves.
To  avoid   over-subtracting  the   sky  background,   the  background
subtraction was  modified from the  pipeline standard as  described by
\citet{applegate14}.

The  FORS2 field-of-view  is  so  small that  only  a few  astrometric
standards are found in the common astrometric reference catalogs. Many
of them are saturated in  our exposures. While we used the overlapping
exposures from all passbands to map them to a common astrometric grid,
the absolute  astrometric calibration was done using  mosaics of F606W
images  centered  on  our  clusters  from  the  complimentary  ACS/HST
programs 12246 (PI Stubbs) and 12477 (PI High).

Because  the   data  were   generally  not  taken   under  photometric
conditions,  the photometric  calibration was  also carried  out using
data  from the  HST  programs.  We derived  a  relation between  F814W
magnitudes and the FORS2 I Bessel filter \citep{chiu16a}
\begin{displaymath}
  m_{\rm I} - m_{\mathrm{F814W}} = -0.052 + 0.0095
  (m_{\mathrm{F606W}} - m_{\mathrm{F814W}})  \;,
\end{displaymath}
from the \citet{pickles98}  stellar library, which is  valid for stars
with $(m_{\mathrm{F606W}} - m_{\mathrm{F814W}}) < 1.7$\,mag. After the
absolute  photometric  calibration of  the  FORS2  $I-$band from  this
relation,  the relative  photometric calibrations  of the  other bands
were  fixed using  a  stellar  locus regression  in  the $(m_{\rm  b},
m_{\mathrm{F606W}}, m_{\rm I}, m_{\rm  z})$ color-space. The inclusion
of F606W data in this process  was necessary because the stellar locus
in $(m_{\rm b}, m_{\rm I}, m_{\rm  z})$ colors has no strong breaks as
in the ($g-r, i-z$) diagrams.  \begin{center}
\begin{table*}
\caption{Optical Imagers Employed in this Study\label{tab:ocameras}}
    \begin{tabular}{lcccccc}
    \hline \hline   

\multicolumn{1}{l}{Site} &
\multicolumn{1}{c}{Telescope} &
\multicolumn{1}{c}{Aperture} &
\multicolumn{1}{c}{Camera} &
\multicolumn{1}{c}{Filters} &
\multicolumn{1}{c}{Field} &
\multicolumn{1}{c}{Pixel scale} \\
\multicolumn{1}{l}{~} &
\multicolumn{1}{c}{~} &
\multicolumn{1}{c}{[m]} &
\multicolumn{1}{c}{~} &
\multicolumn{1}{c}{~} &
\multicolumn{1}{c}{[$\arcmin\times\arcmin$]} &
\multicolumn{1}{c}{[$\arcsec$]}\\
\hline \hline                             

Cerro Tololo & Blanco & 4.0 & MOSAIC-II & $griz$ & $36\times 36$ & $0.27$ \\
Las Campanas & Magellan/Baade & 6.5 & IMACS f/2 & $griz$ & $27\times 27$ & $0.20$ \\
Las Campanas & Magellan/Clay & 6.5 & Megacam & $gri$ & $25\times 25$ & $0.16$ \\
La Silla & 2.2 MPG/ESO & 2.2 & WFI & $BVRI$ & $34\times 33$ & $0.24$ \\
Paranal & VLT Antu & 8.2 & FORS2 & $b_jIz$ & $7\times 7$  & $0.25$  \\ \hline
 \hline
    \end{tabular}
    \end{table*}  
\end{center}

\subsection{Completeness}
\label{sec:completeness}
In the  majority of cases the  photometry is complete to  a $10\sigma$
level  or better  to  a depth  of  $m^*+2$ and  no  correction due  to
incompleteness is necessary.  For the small fraction of the sample for
which this  limit is not  reached, a  correction is applied  to enable
analysis  to   a  common   depth  relative   to  the   cluster  galaxy
characteristic magnitude.  The correction follows our previous work in
\citet{zenteno11}:  We  compare the  $griz$  count  histograms to  the
deeper       Canada-France-Hawaii-Telescope        Legacy       Survey
\citep[CFHTLS,][private     communication]{brimioulle08}\footnote{
  Count histograms correspond  to the D-1 1 sqr.  degree  field, at l=
  $172.0^{\circ}$  and  b =  $-58.0^{\circ}$  with  a magnitude  limit
  beyond r=27  and a seeing  better than 1.0\arcsec} by  dividing both
count histograms.   The resulting curve  is fit by an  error function,
which is  used to account for  the missing objects as  we approach the
$m^*+2$  common depth.   All clusters  covered by  WFI-{\it BVRI}  and
VLT-$Iz$ bands reach  $m^*+2$ to a $10\sigma$ level  and no correction
is applied in those cases.

\section{Cluster Galaxy Populations: tools}

\label{sec:tools}

\citet{song12b} showed  that if the SPT  positional error distribution
is taken into account, BCGs in  the SPT cluster sample are distributed
similarly to  BCGs in  X--ray selected samples.   Furthermore, several
studies have shown the BCG to be  a good proxy for the cluster center,
as  defined  by  X--ray  \citep[e.g.,][]{lin04b,mann12}  and  by  weak
lensing \citep[e.g.,][]{oguri10}, for  the general cluster population.
For the  following analysis we  use the  position of the  observed BCG
\citep[selected within $r_{200}$ following][]{song12b}  as a proxy for
the cluster  center (coordinates listed in  Table~\ref{tab:3}) and its
luminosity as  a limit  on the  bright end,  to reduce  the foreground
contamination.  Error  bars in  variables are estimated  with $\chi^2$
statistics,  where  the  confidence  limits are  defined  as  constant
$\Delta\chi^2$ boundaries \citep{press92}.

\subsection{Radial Distribution of Galaxies}
\label{sec:rp}
While simulations of  dark matter (DM) present a  consistent and clear
picture of  the DM  density profiles  where the  concentration depends
strongly on  redshift but only  weakly on  mass \citep[e.g., $  c(z) =
  5.71^{}_{}\times(1+z)^{-0.47}(M/M_{\rm
    pivot})^{-0.084}$,][]{duffy08},  simulations  of subhaloes,  where
the galaxies are expected to live,  are less clear.  In DM simulations
it  is found  that the  radial  distribution of  subhaloes is  roughly
independent of host  halo mass and redshift.  Also,  as massive haloes
sink more rapidly in the  cluster potential due to dynamical friction,
they    lose   mass    more   rapidly    due   to    tidal   stripping
\citep[e.g.,][]{angulo09}.  When baryon physics is included, the cores
of  the radial  profiles steepen  as  the more  tightly bound  baryons
survive  better  in  the  central   regions  than  DM  only  subhaloes
\citep{nagai05,dolag09}.  These  processes may  have an effect  on the
observed  galaxy  radial   profile  as  well  as   on  the  luminosity
distribution.

On the observational side, no clear redshift trends have been found to
date.  Observations of  the galaxy distribution have  been carried out
in clusters with different redshifts and masses.  For example, using a
local  sample   of  93  groups   and  clusters  with  masses   in  the
$3\times10^{13}$M$_\odot  -  2\times10^{15}$M$_\odot$  range,  and  at
$z<0.06$,  \citet[][hearafter L04]{lin04a}  found  a concentration  of
$c_{\rm  g,200c}=2.9^{+0.21}_{-0.22}$  with  no  evidence  of  a  mass
dependence.   At   a  higher  redshift,   $0.15  \leq  z   \leq  0.4$,
\citet{budzynski12} found $c_{\rm g,200c}\approx 2.6$ independently of
both cluster mass and redshift,  using 55,121 groups and clusters from
the SDSS-DR7.

\citet{muzzin07a}, using 15 CNOC clusters at $0.19 < z < 0.55$, found
a concentration  of $4.13  \pm 0.57$.  At  a much higher  redshift ($z
\approx 1$), \citet{capozzi12}, using 15 clusters with an average mass
of $M_{200}=3.9 \times 10^{14}$M$_\odot$,
found  a concentration  of  $c_{\rm g,200c}=2.8^{+1.0}_{-0.8}$, completely
consistent with the lower redshift cluster samples.

Recently, \citet{vanderburg14,vanderburg15}  studied the  evolution of
the concentration  comparing 60 clusters  at $0.04<  z < 0.26$  and 10
clusters at $0.86< z < 1.34$, finding galaxy density concentrations of
$c_{\rm g,200c}=2.31^{+0.22}_{-0.18}$ (for  the M$^* > 10^{10}M_\odot$
haloes)   and  $c_{\rm   g,200c}=5.14^{+0.54}_{-0.63}$,  respectively.
While  the  low  redshift  sample  agrees  with  the  literature,  the
concentration  found  for the  high  redshift  sample is  higher  than
expected.  As mentioned above, \citet{capozzi12} found a concentration
of $c_{\rm  g,200c}=2.8^{+1.0}_{-0.8}$ at  similar redshifts  but with
masses only twice  as large as \citet{vanderburg15}.   A larger sample
at high  redshift is  needed to  test if this  disagreement is  due to
strong mass dependence  in the concentration of  galaxies in clusters,
or due  to other causes.  With the exception  of \citet{vanderburg14},
these results appear to point to  no evolution in the concentration up
to a redshift  of 1.  We use  the SPT-SZ selected sample  to test this
picture  using  a uniformly  selected  sample  over a  broad  redshift
range. The  radial surface density  profiles are constructed  for both
the  full population  and  the red  population.   The outer  projected
radius ranges from one to three  $r_{200}$, which is the case for most
clusters.  Red  galaxies are  selected if their  colour lies  within a
$\pm 3\sigma(\pm  0.22)$ range  around the  observed red  sequence for
that redshift  \citep[][see \S~\ref{sec:mstar}  for details]{lopez04}.
In a  larger sample it  is possible to measure  the red sequence  as a
function of  redshift, and  then take a  more restrictive  approach to
defining the red sequence population \citep[see][]{hennig16}.

\begin{figure}
\begin{center}
\includegraphics[width=0.49\textwidth]{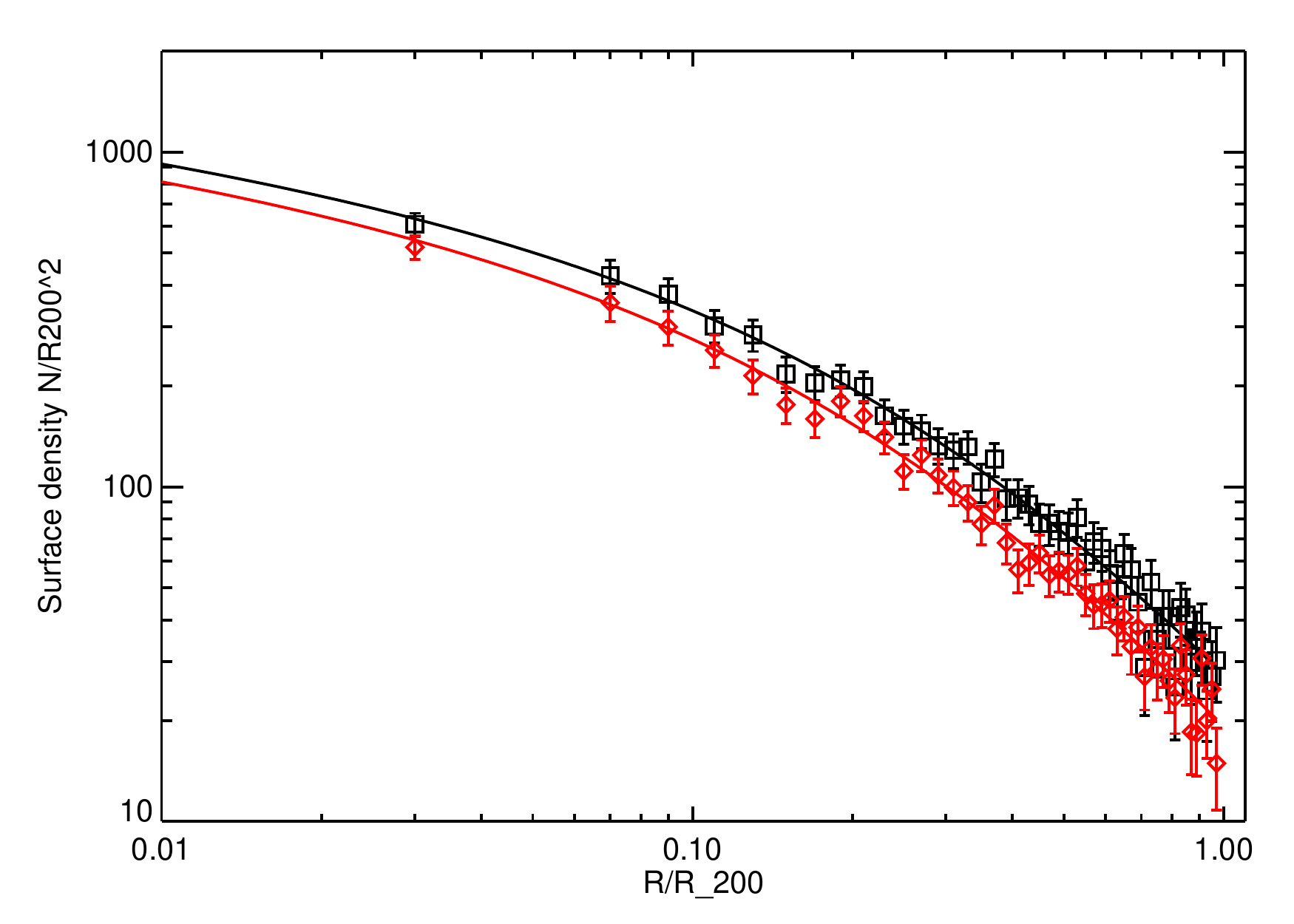}
\caption[Stacked  radial  profiles.]{Radial  profile  of  the  stacked
  sample up to $r_{200}$, using  all galaxies (black) and red sequence
  galaxies (red).   These profiles are  well fit by NFW  profiles with
  the red subsample  somewhat more concentrated than  the full sample,
  with   concentrations   of   \RPredstacksol\   and   \RPallstacksol,
  respectively.
\label{fig:rpall}}
\end{center}
\end{figure}

\begin{figure}
\begin{center}
\includegraphics[width=0.49\textwidth]{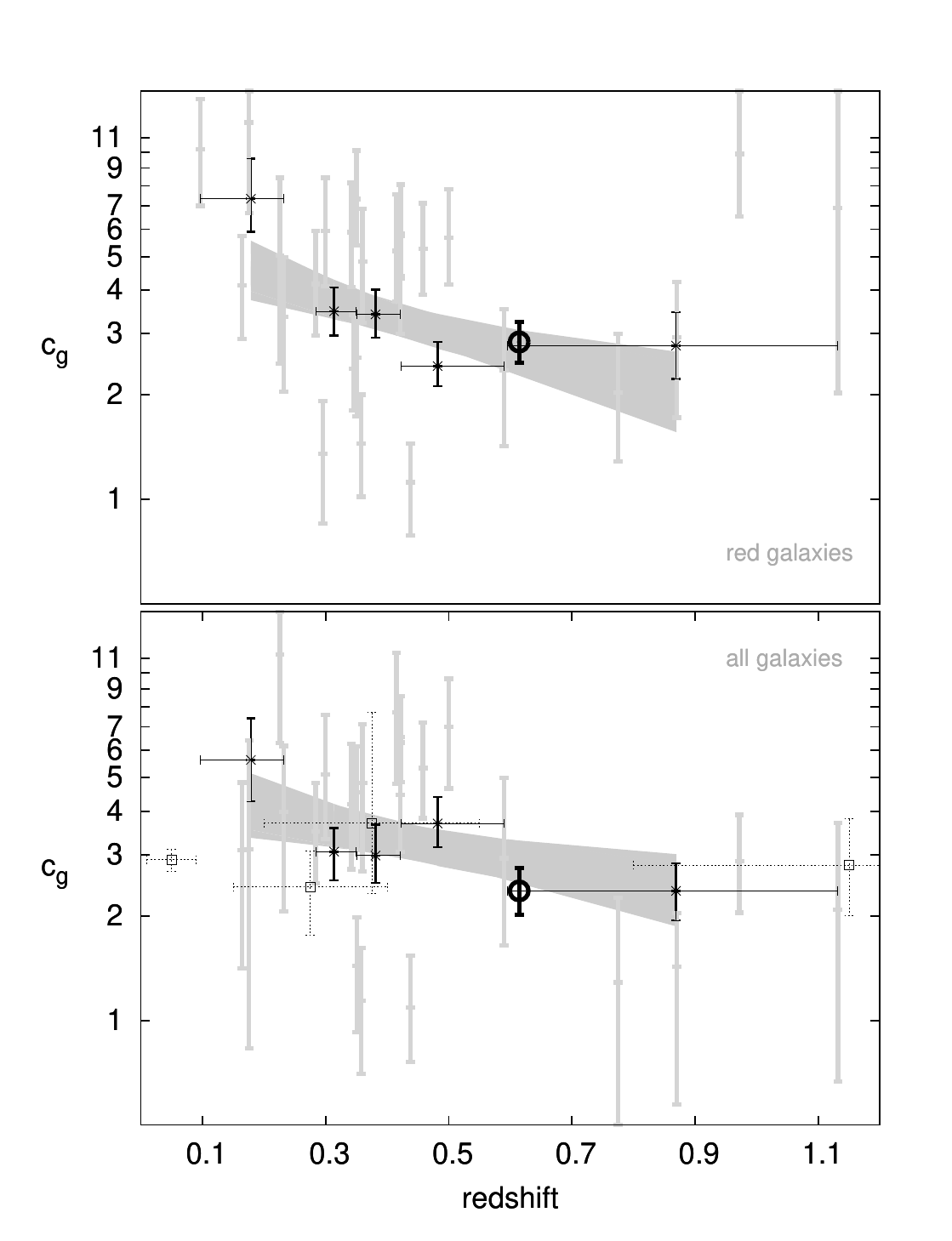}
\caption[Evolution  of   the  concentration   parameter.]{Cluster  NFW
  concentration parameter evolution for red sequence selected galaxies
  (top  panel)  and  all  galaxies  (bottom  panel)  within  projected
  $r_{200}$. Grey  points represent the individual  cluster fits using
  Cash statistics, and the five black points are representative of the
  concentrations  found by  simultaneously fitting  to ensembles  of 5
  clusters  each.    The  central  open  circle   corresponds  to  the
  concentration extracted  from the  fit of the  stacked sample  up to
  $r_{200}$ (see  Fig.~\ref{fig:rpall}).  Open  squares in  the bottom
  panel correspond to  values found in the literature.   There is some
  evidence for redshift evolution in the total sample given a slope of
  \CGallevosolslope,  and for  the red  subsample given  the slope  of
  \CGredevosolslope.  The  apparent trend  is consistent  between both
  the stacked and the individual data.\label{fig:cg_evol} }
\end{center}
\end{figure}

The radial binning is done in two  ways, depending on how the data are
combined and fit.   In one configuration all the data  are stacked and
fitted to  a common  radius $R_{200}$, and  in another  a simultaneous
fitting on subsamples of  individual profiles (multi-fit hereafter) is
performed.  We use  $\chi^2$ statistics (with a number  of members per
bin of $\gtrsim  15$) with a different binning for  each case. For the
multi-fit method,  which involves fitting multiple  individual cluster
radial profiles,  we bin the  data in  $0.05 \times r_{200}$  with the
first bin  and bins  beyond $r_{200}$  being twice  as wide.   For the
stacked case, in which the individual  cluster bins can be much finer,
we use  bins of $0.02  \times r_{200}$ size  with the first  one being
twice as  wide, up to  R$_{200}$.  In addition,  as a cross  check, we
perform individual  fits on single  clusters.  For the  single cluster
fit  we use  bins  of  width $0.02  \times  r_{200}$  size and  beyond
$r_{200}$  double the  width.  As  in the  latter case,  the bins  are
scarcely populated, and we use  \citet{cash79} statistics and a Markov
chain   Monte    Carlo   (MCMC)   Ensemble   sampler    \emcee\   from
\citet{foreman-mackey13}.      The     results    are     shown     in
Fig.~\ref{fig:rpall} and Fig.~\ref{fig:cg_evol}.

As in  \citet{zenteno11}, we  have masked the  saturated stars  in the
field and corrected  for the effective area covered.  This  is done by
gridding the data within a radial bin tangentially by using an angular
bin of 2 degrees (i.e., dividing  the radial bin into 180 tangentially
arranged bins).  Bins that fall within masked areas are discarded from
the  radial area  calculation.  Also,  as  a quality  control, if  two
thirds or more of the area of the annulus is lost, then the annulus is
discarded.  This typically happens at the detector edges.

To compare with previous studies we fit a projected NFW profile to our
radial  distribution.   This  density  is modeled  as  the  number  of
galaxies in  a cylinder within  rings divided  by the ring  area.  The
number  of galaxies  in  a cylinder  of radius  $r$  can be  described
analytically by  integrating the NFW  profile along the line  of sight
\citep[e.g.,][]{bartelmann96}:
\begin{equation}
\label{eq:1}
N_{\rm cyl}(r) = 4\pi \rho_s r_s^3 f(x)
\end{equation}
\begin{displaymath}
f(x) =
\begin{cases}
  \ln\frac{x}{2} + \frac{2}{\sqrt{x^2-1}}{\rm arctan}\sqrt{\frac{x-1}{x+1}} & \text{if } x > 1, \\
  \ln\frac{x}{2} + \frac{2}{\sqrt{1-x^2}}{\rm arctanh}\sqrt{\frac{1-x}{x+1}} & \text{if } x < 1\\
  \ln\frac{x}{2} + 1 & \text{if } x = 1
\end{cases}
\end{displaymath}
where $\rho_{\rm  s}$ is the central  density, $r_s=r_{200}/c_{\rm g}$
is  the scale  radius, $c_{\rm  g}$  is the  galaxy concentration  and
$x=c_{\rm g}r/r_{200}$.  We can parametrize  this as a function of the
number of galaxies within a cylinder of $r_{200}$ radius:
\begin{displaymath}
N_{\rm cyl}^{r_{200}} = 4\pi \rho_{\rm s} r_{\rm s}^3 f(c_{\rm g}).
\end{displaymath}
Combining this with  Eq.~\ref{eq:1} we can write  the projected number
of galaxies within $r_{200}$ as a function of $N_{\rm cyl}^{r_{200}}$:
\begin{equation}
\label{eq:2}
N_{\rm cyl}(r) =N_{\rm cyl}^{r_{200}} \frac{f(x)} {f(c_{\rm g})}
\end{equation}
Thus, in the end we fit $c_{\rm g}$, $N_{\rm cyl}^{r_{200}}(M)$ plus a
flat background  $N_{\rm bkg}$  to our  data.  Note  that even  if all
cluster galaxy distributions had the same shape, we would still expect
the   number   of   galaxies   within  the   virial   region   $N_{\rm
  cyl}^{r_{200}}(M)$ to exhibit a cluster mass dependence.\\

Due to the heterogeneity of our optical imaging dataset we have radial
profiles  extending from  one  to  several $r_{200}$,  and  it is  not
possible  to  define  a  region  for  background  estimation  that  is
uncontaminated by the cluster.  We  approach this problem in two ways:
(1) we simply discard the  background information and combine the data
over the region where all clusters have coverage ($\sim 1r_{200}$, see
Fig.~\ref{fig:rpall})  and  (2)  we simultaneously  fit  all  clusters
making use of the common NFW shape parameters while marginalizing over
individual  cluster backgrounds.   That  is, we  fit  each cluster  by
fixing  a   common  $c_{\rm   g}$  and  $N_{\rm   cyl}^{r_{200}}$  but
marginalizing over  the individual  cluster background  $N_{\rm bkg}$.
While  in the  former case  the  $\chi^2_{\rm stack}$  comes from  the
single  fit,  in  the  latter,   the  stack  $\chi^2_{\rm  stack}$  is
calculated   as  the   sum  of   the  individual   cluster  $\chi^2_i$
contributions.  Errors are reported as the projection of the 1$\sigma$
contour      for       1      parameter      \citep[$\Delta\chi^2_{\rm
    stack}=1$;][]{press92}    for    $c_{\rm     g}$    and    $N_{\rm
  cyl}^{r_{200}}$.

Although the mass range in the  current sample is small there are mass
dependencies  which need  to  be  accounted for  in  the stacking  and
multi-fit processes.   We do this by  varying $N_{cyl}^{r_{200}}$ from
Eq.~\ref{eq:2} as a function of the  cluster mass $M$ in the following
way:
\begin{displaymath}
N_{\rm cyl}^{r_{200}}(M)=N_{\rm cyl,piv}^{r_{200}}\left[\frac{M}{M_{\rm piv}}\right]^{\gamma}
\end{displaymath}
where $\gamma=0.87$ (L04)  and the pivotal mass is $M_{\rm piv}=10^{15}$M$_\odot$. 

\subsection{Luminosity Function}
\label{sec:lf}
As galaxy clusters grow by accreting galaxies from the cosmic web over
time, these galaxies are also transformed by processes such as merging
and ram  pressure stripping, formation of  new stars and the  aging of
their                        stellar                       populations
\citep[e.g.,][]{dressler80,butcher84,lin04a,lopes14,gandhi13}.     The
evolution of the cluster luminosity function encodes information about
these  physical processes  and is  therefore an  important tool.   For
example,  by studying  the  bright end  of the  cluster  LF, which  is
dominated by luminous early-type  galaxies, several studies have shown
that the  evolution is  consistent with  a passively  evolving stellar
population     \citep[e.g.,][]{depropris99,lin06,andreon08,capozzi12},
independent  of  the  cluster's dynamical  state  \citep{depropris13}.
This indicates that the cluster  galaxies have their stellar component
in place at high redshift \citep[$z\gtrsim 2-3$; e.g.,][]{mancone10}.

The  individual  cluster LF  is  constructed  using sources  within  a
projected $r_{200}$,  centered on the  BCG.  We perform  a statistical
background subtraction  using a  background region  at $r>1.5r_{200}$.
In general, we make use of the  photometry up to a 10$\sigma$ level at
an $m^*+2$ depth or  even deeper.  The projected, background-corrected
LF is then  de-projected using an NFW profile with  a concentration of
\LFCGCORRECTIONALL~ and  \LFCGCORRECTIONRED, which corresponds  to the
stack  value  in  Fig.~\ref{fig:rpall}  for   the  full  and  the  red
populations, respectively.   Finally, the  cluster LF is  divided into
the different magnitude bins and scaled by the cluster volume in Mpc.

Corrections due to masked  regions and background over-subtraction are
applied here as well.  In the  case of masked regions within $r_{200}$
we correct for the missing cluster galaxies using the NFW profile with
the  concentration $c_{\rm  corr}$.  Also,  using the  same model,  we
correct for the over subtraction due to cluster galaxies contaminating
the  background  dominated  region.   This  over  subtraction  can  be
expressed  by  an extra  term  $N_{\rm  clus,true}^{>1.5r200}$ in  the
background:
\begin{equation}
\label{eq:ncorrection}
N_{\rm clus,obs}^{r200}=N_{\rm total}^{r200}-A_{\rm N}\times(N_{\rm back}^{>1.5r200}+N_{\rm clus,true}^{>1.5r200})
\end{equation}
where $A_N$ is the area  normalization between cluster and background.
Under the assumption that there  is no luminosity segregation and that
the galaxy distribution is well described by an NFW model with a given
concentration, we  can connect  the over  subtraction to  the galaxies
within  $r_{200}$  as  $N_{c\rm  lus,true}^{>1.5r200}=\tau(c_{\rm  g})
N_{\rm clus,true}^{r200}$.  Combining with Eq.~\ref{eq:ncorrection} we
have a correction:
\begin{displaymath}
N_{\rm clus,true}^{r200} = \frac{N_{\rm clus,obs}^{r200}}{(1 -A_{\rm N}\times \tau(c_{\rm g}))} = C \times N_{\rm clus,obs}^{r200}.
\end{displaymath}
The average correction C is of the order of ~1.11.  

Finally, two of  the clusters have only imaging from  VLT/FORS2 with a
FOV  of 7\arcmin  $\times$7\arcmin, covering  less than  $1.5r_{200}$.
For SPT-CLJ2106-5844 at z=1.131, the  background area is re-defined as
the area at $r > r_{200}$  with a corresponding correction C, of 1.49.
For  the cluster  SPT-CLJ0102-4915,  this re-defined  area  is at  the
detector edge  and an  external background is  used.  As  a background
area     we    use     the    {\it     Cosmic    Evolution     Survey}
\citep[COSMOS,][]{scoville07a} data, avoiding regions with known large
scale       structures       at       the       cluster       redshift
\citep{scoville07b}\footnote{
  $149.4^\circ  \le {\rm  R.A.}  \le  150.2^\circ$ and  $1.5^\circ \le
  {\rm decl.} \le 2.2^\circ$}.\\

Once the  LF is  constructed we  fit it  by the  three parameter
Schechter function (SF) \citep{schechter76},
\begin{displaymath}
\phi(m)=0.4 \ln(10)\ \phi^*10^{0.4(m^*-m)(\alpha+1)}\exp(-10^{0.4(m^*-m)}).
\end{displaymath}
We  fit  the  SF  to  the stack,  and  to  the  individual  luminosity
functions.  In the single cluster case, simulations show that there is
little  constraint on  \mstar~if the  three variables  are allowed  to
float within our typical luminosity  range (see \S \ref{sec:sims}), so
our  approach  is to  extract  the  parameters $\phi^*$,  \mstar,  and
$\alpha$ by fixing  one parameter and leaving the other  two to float.
Specifically, for the \mstar~evolution  analysis, we fix $\alpha$.  We
note  that  the  three  parameters   of  the  Schechter  function  are
correlated, so  fixing one variable  to the  wrong value will  have an
impact on the free parameter.

For the stacked LF we fit all  three parameters.  We bring the data to
a common  frame fitting in the  space of $m-m^*_{\rm model}$,  using a
Composite  Stellar   Population  model  (see   \S~\ref{sec:mstar}  for
details).  Once  the data are brought  to this common frame,  they are
stacked using an inverse variance weighted average:
\begin{equation}
\label{eq:stack}
N_{j}=\frac{\sum_i  N_{ij}^{z=0} / \sigma_{ij}^2}{\sum_i \sigma_{ij}^2}
\end{equation}
where  $N_{ij}^{z=0}$  is  the  number  of  galaxies  per  volume  per
magnitude  at redshift  zero, in  the $j$th  bin corresponding  to the
$i$th cluster's LF and $\sigma_{ij}$  is the statistical poisson error
associated.   We  obtain  $N_{ij}^{z=0}$   by  correcting  it  by  the
evolutionary    factor    $E^2(z)$,   where    $E(z)=\sqrt{\Omega_{\rm
    m}(1+z)^3+\Omega_\Lambda}$.   This  scaling   is  appropriate  for
self-similar  evolution where  the characteristic  density within  the
cluster  virial region  will scale  with the  critical density  of the
universe.

The errors of the stacked profile are computed as
\begin{displaymath}
\delta N_{j}=\frac{1}{(\sum_i \sigma_{ij}^2)^{1/2}}
\end{displaymath}
We adopt $\alpha$  from the stacked LF for the  evolution study of the
single cluster characteristic magnitudes \mstar.

\subsection{Composite Stellar Population Models}
\label{sec:CSP}
Several  studies  have  shown  that  \mstar\  evolution  can  be  well
described by a  passively evolving stellar population  that has formed
at                            high                            redshift
\citep[e.g.,][]{depropris99,andreon06a,lin06,depropris07,mancone10}.
Empirically, these  Simple Stellar  Population (SSP) models  have been
used to  predict red sequence  colors that  are then used  to estimate
cluster  redshifts   with  characteristic  uncertainties   of  $\delta
z\sim0.025$ \citep[e.g.,][]{song12a,song12b}.   Generally speaking, in
an analysis of cluster galaxy  populations over a broad redshift range
it  is  helpful  to  have  a model  within  which  the  evolution  and
$k$-corrections  are   self-consistently  included  to   simplify  the
comparison of  cluster populations  at different redshifts  within the
observed bands.

In this  analysis we create  red sequence  CSP models for  Mosaic2 and
IMACS   $griz$,  WFI   $BVRI$,   and  VLT   $BIz$   bands  using   the
\citet{bruzual03}   SSP  models   and  the   EzGal  python   interface
\citep{mancone12a}.  The  models consist  of an  exponentially falling
star formation rate with a decay time  of 0.4 Gyr, Salpeter IMF, and a
formation redshift of 3.  We  use in total six different metallicities
to   introduce  the   tilt   in  galaxy   red   sequence  within   the
color-magnitude  space.   To  calibrate  these  models  we  adopt  the
measured  metallicity-luminosity relation  for  Coma cluster  galaxies
\citep{poggianti01b,mobasher03}.   This  procedure   then  requires  a
further     adjustment     of     the    Coma     $L^*$     luminosity
\citep[][]{iglesias-paramo03}  brightening  it  by 0.2  magnitudes  to
reproduce the  observed colour of  the Coma cluster.   This calibrated
set of  CSP models allows  us to  predict the apparent  magnitudes and
colors of  all our  cluster populations within  the range  of relevant
observed bands.  As described  in \S~\ref{sec:results} below, by using
the full sample of clusters we can  test whether this set of models is
consistent with the real galaxy populations.
\subsection{Simulated Galaxy Catalogs}
\label{sec:sims}

To  test our  methods, find  the best  stacking strategy  and quantify
possible  biases, we  create simulated  galaxy catalogs  of a  typical
cluster.  We re-create  a galaxy cluster using the  number of galaxies
in a  cluster of mass  $M_{200} = 1.3\times 10^{15}  $M$_\odot$, given
the  expected  number  of  galaxies  from  measurements  of  the  halo
occupation number (HON at low  redshift, L04) and with a concentration
of 3 over a typical angular region  on the sky.  This corresponds to a
spherical number of  galaxies, within $r_{200}$ and up  to a magnitude
of $m^*+3$,  of $N_{\rm sph}^{r_{200}}=  335$ and its  projected value
$N_{\rm cyl}^{r_{200}}=  443$.  Although $m^*+2$ is  our typical depth
we  extend the  cluster counts  to  $m^*+5$ for  testing purposes.  No
luminosity  segregation is  included. We  assign galaxy  magnitudes to
match  an LF  with  $\alpha=-1.2$  and $m^*=m^*_{\rm  model}(z=0.35)$,
while  $\phi^*$ is  set  by $N_{\rm  sph}^{r_{200}}$.   The number  of
background galaxies used corresponds to  45,000 sources in the $m^*-3$
to  $m^*+5.5$   luminosity  range   with  a   brightness  distribution
equivalent   of  the   CFHTLS   $r-$band  count   histogram  used   in
\S~\ref{sec:completeness}.   The construction  of the  radial profiles
and luminosity functions is done using  the same tools as for the real
clusters, accounting for  the masked areas due to CCD  gaps, stars and
missing CCDs.

As we mention in \S~\ref{sec:rp},  the multi-fit stack approach uses a
typical bin  size of $0.05r_{200}$, while  the first bin and  the bins
beyond $r_{200}$ are  twice as wide.  This configuration  is chosen to
balance a good number of galaxies ($\gtrsim 15$) per bin with the need
to have  narrow enough bins to  be able to constrain  $c_{\rm g}$.  We
fit for $c_{\rm g}$ and $N_{\rm cyl}^{r200}$ and marginalize over each
individual cluster background.  We demonstrate this with the multi-fit
method on  five clusters using  the region extending up  to $3r_{200}$
over 20 realizations, the  concentration is recovered within 1$\sigma$
(\onessims).

Another way  to use  the data  is to stack  the cluster  data up  to a
common maximum radius.   In this case there are more  galaxies per bin
than in the single cluster case, giving us the chance to explore finer
bins and to test that our results are not biased due to the chosen bin
size.  The common maximum radius is  reached at $\sim r_{200}$, set by
the  lowest redshift  cluster. We  use  a bin  set of  0.04, 0.02  and
0.1$r_{200}$ for the first bin, the  bins below $r_{200}$ and the bins
at $>r_{200}$, respectively.  Simulations show  that in the case of 25
clusters in  the stack,  the input  concentration is  recovered within
1.5$\sigma$  ($3.62^{+0.48}_{-0.41}$).  In  comparison, when  the same
data are stacked up to 3$r_{200}$, the input values are recovered well
within 1$\sigma$.

Using  the multi-fit  stack binning  configuration, we  also test  the
individual results.  Fitting for  the radial profile parameter $c_{\rm
  g}$,  $N_{\rm  cyl}$ and  background  in  each individual  simulated
cluster,  over  the  100  realizations,   the  weighted  mean  of  the
concentration is  recovered well within 1$\sigma$  ($c_{\rm g}=2.97\pm
0.12$).  These tests give us  confidence that our binning strategy and
our  scripts are  suited for  use  in extracting  measurements of  the
concentration of  the galaxy clusters  in this study with  biases that
are at or below the statistical uncertainty.

In  the  case  of  the  luminosity  function, we  use  and  apply  the
configuration  and corrections described  in \S~\ref{sec:lf}  (0.5 mag
bin, count  correction due to background over subtraction, star-masked
areas, CCD  gaps, etc.)  to test  our scripts and  assess the
level of bias and or scatter under this configuration.

Simulations  demonstrate  that  simultaneously fitting  all  three  SF
parameters provides  only weak constraints  on \mstar, given  that the
typical depth pushes  to $m^*+2$.  To overcome this we  fix one of the
three parameter and explore the other  two: when $\alpha$ is fixed the
weighted  mean  value recovered  for  \mstar\  is within  1.6$\sigma$.
Conversely, if \mstar~is  fixed, $\alpha$ is recovered  well to within
1$\sigma$.  In  the case of the  HON, when \mstar\ is  fixed, the true
HON is  recovered to 0.6$\sigma$  and to 3.2$\sigma$ when  $\alpha$ is
the variable fixed  to the input value.  Accordingly  our first choice
is to fix \mstar\ when studying the HON.

\section{Results}
\label{sec:results}
\subsection{Radial Profile}

The composite profiles for the full and red sequence selected galaxies
in the full sample of clusters are shown in Fig.~\ref{fig:rpall}.  The
lines  trace out  the  best fit  NFW profiles,  which  provide a  good
description of  the stacked galaxy  profiles in both cases.   The best
fit concentration for  the red galaxy sample  is \RPredstacksol, which
is   somewhat  higher   than  that   for  the   total  population   of
\RPallstacksol.   The higher  concentration for  the red  subsample is
consistent with the radial variations of red fraction found in optical
studies           of          other           cluster          samples
\citep[e.g.,][]{goto04,verdugo12,ribeiro13,gruen13a}.

Our  measured concentration  for  the full  sample \RPallstacksol\  is
somewhat    lower     when    compared    to     previous    estimates
$2.9^{+0.21}_{-0.22}$ at  redshift zero (L04)  and $2.8^{+1.0}_{-0.8}$
at  $z\sim 1$  \citep[][]{capozzi12}.  Given the  high  masses of  our
sample, one may wonder if the differences reflect a mass dependence on
the concentration.   While in DM  simulations more massive  halos have
lower concentration,  the same  simulations do not  show such  a trend
with  galaxies.  Some  analyses  have   shown  a  steep  inverse  mass
dependence with  concentration \citep{hansen05}, while  other analyses
\citep[including many  of the  same clusters][]{budzynski12}  found no
such trend.  They attribute the  difference to different approaches in
defining the radius in the  two studies. \citet{vanderburg15} did find
a steep mass-concentration relation,  although the two cluster samples
are at  very different  redshifts.  Nevertheless,  for the  high mass,
low-redshift sample,  the concentration found  by \citet{vanderburg15}
of  $c_{\rm g,200c}=2.31^{+0.22}_{-0.18}$  is  in excellent  agreement
with ours.  \citet{hennig16} used an  SPT selected sample with a lower
mass  average  finding  higher  concentrations  of  \hennigcgall\  and
\hennigcgred\   for  the   total  and   the  red   galaxy  population,
respectively.    This   overall   picture   seems  to   point   to   a
mass-concentration relation steeper than DM only simulations.

The  concentration measured  as a  function  of redshift  for the  SPT
sample is shown in Fig.~\ref{fig:cg_evol}. The individual cluster fits
are  shown in  light grey,  pointing  to an  apparent evolution.   The
multi-fit over five bins with five  clusters in each bin confirms this
picture.  Fitting  a slope and  intercept to  the red sample  and full
subsample we find \CGredevosol\  and \CGallevosol\ which correspond to
\CGredevosignificance   $\sigma$   and   \CGallevosignificance$\sigma$
significance respectively,  of a possible evolution.  Also, the result
from  the stack  over all  redshifts is  consistent with  this formula
within the errors, as expected.

\begin{figure}
\begin{center}
\includegraphics[width=0.49\textwidth]{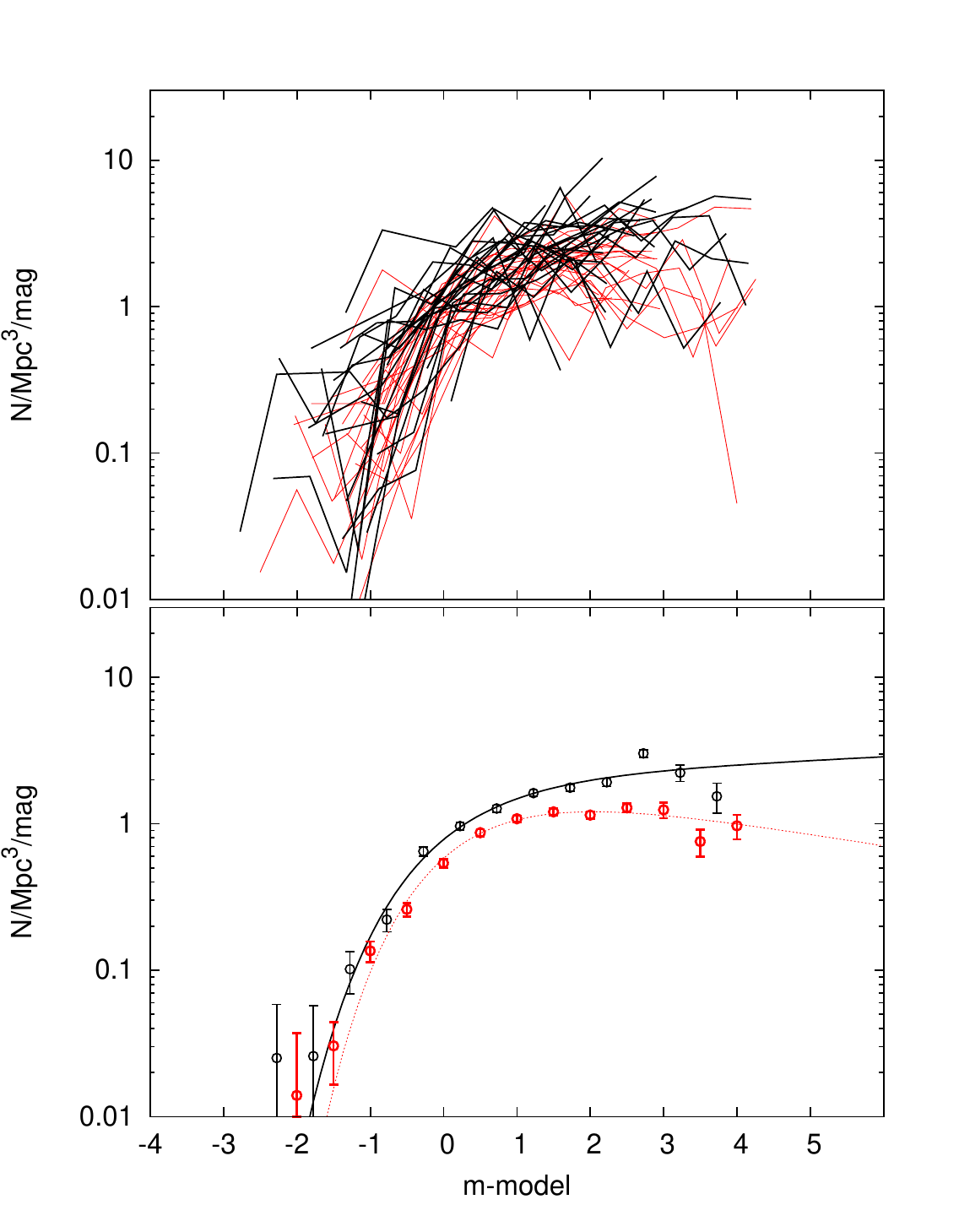}
\caption[Individual and  stacked luminosity function.]{ We  plot 24 of
  the  26  individual  LFs  (top) versus  $m-m^*_{\rm  model}$,  where
  $m^*_{\rm model}$ is the  predicted CSP characteristic luminosity at
  the redshift of the cluster.   Each individual LF is extracted using
  the band  redward of the  4000\AA~break.  The two  excluded clusters
  included  the  lowest  redshift  system where  our  imaging  is  not
  adequate and another system that has a foreground star field, making
  it difficult to identify the  faint galaxy population.  The BCGs are
  excluded.  The weighted averaged  luminosity function appears below.
  In black the total population is  shown, and in red the red-sequence
  population  is  displayed.   Bins  with at  least  two  contributing
  clusters  are  shown.  The  fit  for  the  all galaxies  stacked  is
  \phiallsol\   and   \alphaallsol\   (\chiallsol).    The   fit   for
  red-sequence galaxies is \phiredsol\ and \alpharedsol\ (\chiredsol).
  \label{fig:lf_stack}}
\end{center}
\end{figure}

\subsection{Luminosity Function}
Several studies have found that the steepness of the faint end depends
on the band chosen \citep[e.g.,][]{goto02,goto05}, as bands bluer than
the  4000\AA~break  are more  sensitive  to  younger populations.   We
systematically select the nearest band redward of the 4000\AA, and are
therefore less sensitive  in our study to recent  star formation.  The
bands were chosen as follows: $r$-band for $0 < z \leq 0.35$, $i$-band
for $0.35 < z  \leq 0.70$, $z$-band for $ z > 0.70  $.  In the case of
{\it BVRI} the  conditions were $V$-band for $0 <  z < 0.20$, $R$-band
for  $0.20 <  z <  0.40$ and  $I$-band for  $z >  0.40$.  For  the two
clusters with VLT data ($z \geq 0.7$), $z_{Gunn}$ was used.

\subsubsection{Stacked Luminosity Function}
\label{sec:stackedlf}

For the stacked LF we use  24 clusters.  The two excluded clusters are
SPT-CL   J2201-5956   which,   with   a   $z=0.098$   and   $1.5\times
10^{15}$M$_\odot$  mass,  has a  projected  $r_{200}$  outside of  the
field-of-view, making it all but impossible to estimate the background
contribution,  and  SPT-CLJ0555-6406, which  has  a  star field  as  a
foreground that makes the cluster normalization unreliable.

As we mentioned  in \S~\ref{sec:sims}, fitting all  three variables in
the LF produces large errors in the parameter exploration.  To address
this  problem, we  use \mstar\  from the  model or  $\alpha$ from  the
stacked LF  to explore the remaining  two LF parameters.  In  spite of
the large errors during three parameter  SF fits, we need at a minimum
to check that  the \mstar\ evolution is consistent  with our passively
evolving  CSP model.   Doing this  we find  that a  linear fit  to the
observed $m^*$ distribution as a function of redshift has a zero point
of \mstarevointerallfree\  and a  slope of  \mstarevoslopeallfree\ for
the  total population.   That is,  the normalization  of our  model is
consistent to within the uncertainties  with the data, and the dataset
over  this  broad  range  of  redshifts provides  no  evidence  for  a
deviation  from the  model.  We also  compare our  model  to red  only
galaxies finding  a slope of \mstarevoslopeallfreered,  also providing
no evidence of evolution of $m^*$ beyond the model.  Nevertheless, the
zero  point found  is \mstarevointerallfreered\  which is  significant
enough to warrant  further model adjustments to account  for the known
covariance between $\alpha$  and $m^*$. We explore  corrections in the
model for the red-only population in \S~\ref{sec:mstar}.

We  proceed to  stack the  LF  using the  model \mstar\  to bring  all
clusters to the same relative  reference frame of $m-m^*_{\rm model}$.
Next, we combine the data using  the weighted average in each bin (see
Eq.~\ref{eq:stack}).  The stacked  LF, as well as  the individual LFs,
for all  and red galaxies  are shown in  Fig.~\ref{fig:lf_stack}. Data
points shown  contain contributions from  at least two  clusters.  The
fit to  the stacked  LF yields \phiallsol\  and \alphaallsol\  for the
total population,  and \phiredsol\ and \alpharedsol\  for red sequence
galaxies.

Our  best fit  faint  end  $\alpha$ for  these  SZE selected  clusters
spanning a  large range  of redshift  is consistent  with measurements
using variously selected samples at different redshifts \citep[][which
  provided     measurements     of    $\alpha=-1.09     \pm     0.08$,
  $-1.11^{+0.09}_{-0,07}$,    $-1.01^{+0.09}_{-0,07}$,    $-1$,    and
  $-1.05\pm0.13$,
  respectively]{gaidos97,paolillo01,piranomonte01,barkhouse07,popessoII05}.

Initially $\phi^*$  seems lower  than in L04,  a previous  study.  L04
found  a best  fit for  their data  of $\phi^*=4.43\pm0.11  ~h^3_{70}~
\rm{Mpc}^{-3}$ for  $\alpha=-0.84 \pm  0.02$ (best  fit), but  found a
lower $\phi^*=3.00\pm0.04  ~h^3_{70}~ \rm{Mpc}^{-3}$ when  $\alpha$ is
fixed to$  = -1.1$, noting  that both $\alpha$'s described  well their
data.  As  our systems are  more massive and the  slope of the  HON is
less than  unity it is  expected that  our $\phi^*$ solution  would be
lower than  that measured  for lower mass  systems.  L04  also explore
this possibility, using their 25  most massive systems, with mean mass
of M$_{500} = 5.3\times  10^{14}\ $M$_\odot$ finding $\alpha=-0.84 \pm
0.03$  and $\phi^*=4.00\pm0.16  ~h^3_{70}~ \rm{Mpc}^{-3}$.   Given the
dependence of  $\alpha$ and  $\phi^*$ shown and  the mass  range, this
result using a  redshift zero sample of clusters  and 2MASS photometry
seems to be  consistent with our result.  A larger  cluster mass range
is needed to carry out a more precise test.

\begin{figure}
\begin{center}
\includegraphics[width=0.44\textwidth]{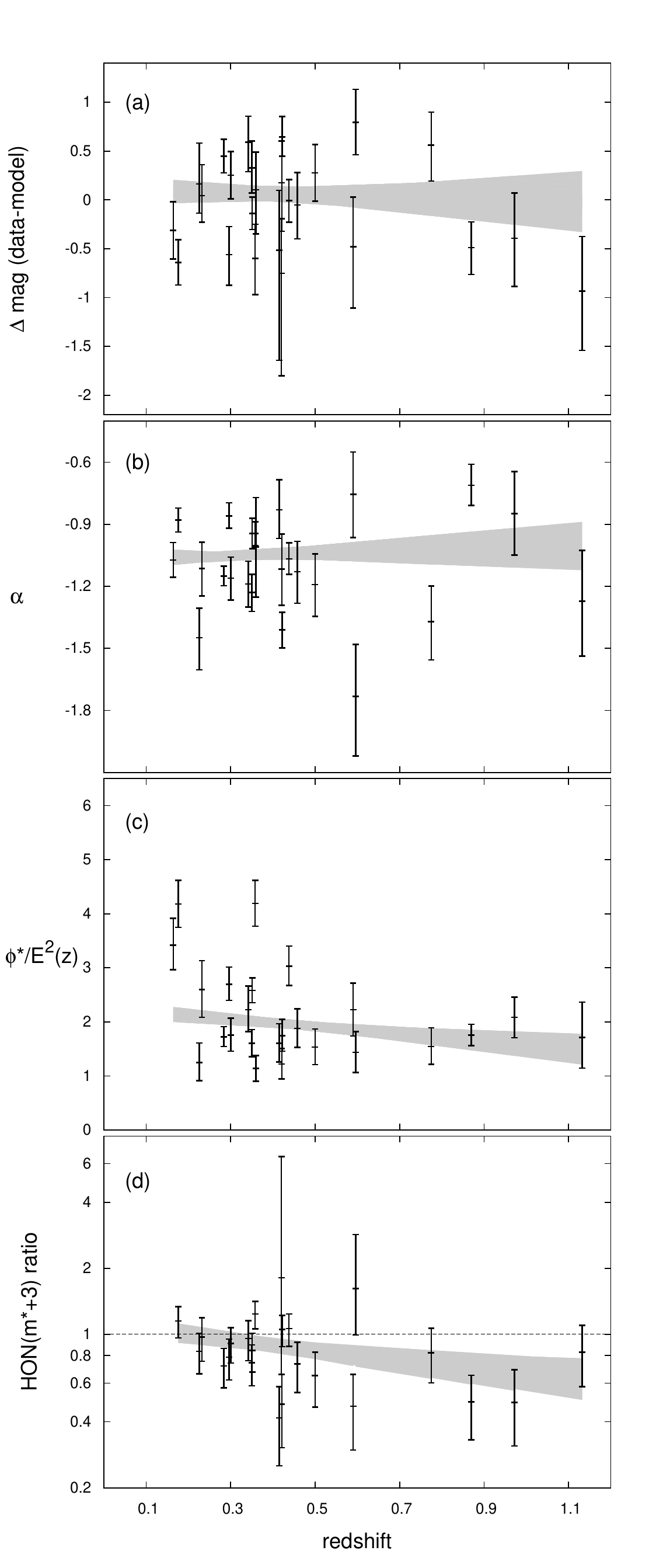}
\caption[Evolution   of   the  luminosity   function   parameters.]{LF
  parameter evolution  with redshift.   As noted  before, the  LFs are
  extracted using  the band  redward of the  4000\AA~break.  We  fit a
  line in each case, marking the allowed 1$\sigma$ region.  {\it Panel
    (a):} There  is no significant evolution  in $\Delta mag=(m^*_{\rm
    model}-  m^*$),   indicating  the   CSP  model  provides   a  good
  description  of cluster  galaxies  over this  redshift range.   {\it
    Panel (b):}  Evolution of $\alpha$  is suggested by the  data with
  best   fit  line   having   intercept   \alphaevointer\  and   slope
  \alphaevoslope.   {\it Panel  (c):}  $\phi^*/E^2(z)$ extracted  when
  fixed    \mstar\    is    consistent   with    no    evolution    at
  \phiallsignificancemfix $\sigma$  level.  {\it Panel (d):}  Ratio of
  HON  from this  work and  the redshift  independent L04  prediction.
  Slope   and   intercept   are   found  to   be   \HONevoslope\   and
  \HONevointer\  at 1  $\sigma$  respectively, which  indicate a  mild
  evolution where  $z=1$ clusters  have typically 30\%  fewer galaxies
  than their low redshift counterparts of the same mass. 
   \label{fig:evolution}}
\end{center}
\end{figure}

\begin{figure}
\begin{center}
\includegraphics[width=0.44\textwidth]{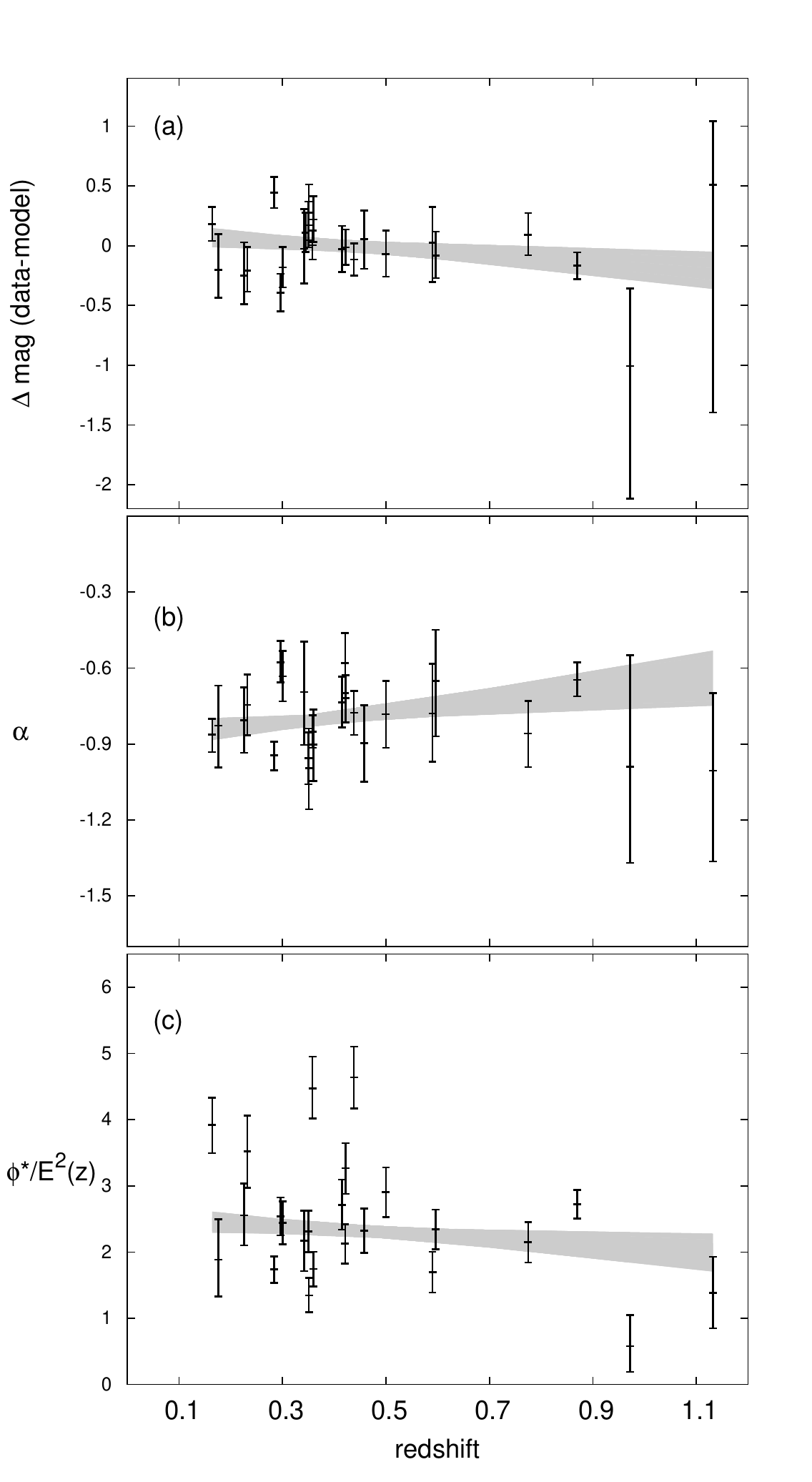}
\caption[Evolution  of   red  luminosity   function  parameters.]{Same
  as~Fig.\ref{fig:evolution}  for red  sequence galaxies.   {\it Panel
    (a):} There  is no significant evolution  in $\Delta mag=(m^*_{\rm
    model}  -  m^*)$,  indicating  the   CSP  model  provides  a  good
  description  of cluster  galaxies  over this  redshift range.   {\it
    Panel (b):}  Evolution of $\alpha$  is suggested by the  data with
  best  fit   line  having  intercept  \alpharedevointer\   and  slope
  \alpharedevoslope.  {\it Panel  (c):} $\phi^*/E^2(z)$ extracted when
  fixed    \mstar\    is    consistent   with    no    evolution    at
  \phiredsignificancemfix $\sigma$ level.
   \label{fig:evolution_red}}
\end{center}
\end{figure}

\subsubsection{Evolution of \mstar}
\label{sec:mstar}
Several previous studies have shown  that the evolution of \mstar\ for
cluster galaxy  populations can be  described by a  passively evolving
stellar       population      formed       at      high       redshift
\citep[e.g.,][]{depropris99,andreon06a,depropris07,suhada12,stalder13}.
We test  this result by fitting  the LF using \mstar\  and $\phi^*$ as
free  parameters while  fixing $\alpha$  to the  measurement from  the
stack.   We compare  the  obtained  \mstar\ to  a  CSP  model that  is
produced as described in Sec.~\ref{sec:CSP} above.

In panel (a) of Fig.~\ref{fig:evolution}  we show a comparison between
the observed  \mstar\ and  our CSP  model.  From  this figure  and the
1$\sigma$ (grey) area, it is clear that  the data and our CSP model is
in  good overall  agreement.  A  linear  fit with  redshift yields  an
intercept of \mstarevointer\ and a  slope of \mstarevoslope. Thus, our
CSP model  of an  exponential burst  of star formation  at z=3  with a
decay  time of  0.4~Gyr  and a  Salpeter  IMF tuned  with  a range  of
metallicities to reproduce the tilt  of the red sequence population at
low  redshift provides  a good  description  of the  evolution of  the
cluster  galaxy populations  over a  broad range  of redshift.   It is
important to  emphasize that our  $m^*$'s are extracted from  the band
that is  just redward  of the  4000~\AA\ break, a  band that  would be
expected to  be relatively insensitive  to recent star  formation.  If
red sequence galaxies  are used a similar result is  obtained. The top
panel  of  Fig.~\ref{fig:evolution_red}  shows  \mstar\  not  evolving
within   the    sample   redshift   range   (the    slope   found   is
\mstarredevoslope).  While there is no evidence for evolution of $m^*$
for the red population, a  non-zero weighted average overall offset of
0.46    is   found    (and    applied   to    the    panel   (a)    of
Fig.~\ref{fig:evolution_red}).   We attribute  this difference  to the
$m^*-\alpha$  covariance and  we  apply this  correction  for the  red
galaxies only model by dimming  the models by the corresponding value.
This correction in the model  normalization is important, as by fixing
a  wrong  $m^*$  model  we  would infer,  for  example,  an  incorrect
$\alpha$.   As   a  sanity  check   we  remind  the  reader   that  in
\S~\ref{sec:stackedlf}      we     found      an     intercept      of
\mstarevointerallfreered\ for  a three parameters SF  fitting, in full
agreement with the correction described above.

\subsubsection{Evolution of $\phi^*$}
\label{sec:phi}
The LF normalization ($\phi^*$) is  the number of galaxies per Mpc$^3$
per unit magnitude, and it informs us, once the universal evolution of
the critical  density is scaled  out, about possible evolution  of the
number  density  of  galaxies  near the  characteristic  magnitude  in
cluster environment.  In  our study we are using the  SZE data to give
us  the cluster  mass $M_{200}$,  the mass  within the  region of  the
cluster that  has a mean  density of  200 times the  critical density.
Because  the  critical density  evolves  with  redshift as  $\rho_{\rm
  crit}\propto E^2(z)$ where $H(z)=H_0E(z)$, we expect to see a higher
characteristic galaxy density at high redshifts.  Thus, to explore for
density   evolution   beyond   this   we   examine   measurements   of
$\phi^*/E(z)^2$ in  the case  where $\alpha$ is  a free  parameter and
$m^*$  comes from  the  CSP model.   Results appear  in  panel (c)  of
Fig.~\ref{fig:evolution}       for       all       galaxies,       and
Fig.~\ref{fig:evolution_red}  for  the  red sequence  sub-sample.   By
fitting  a  linear  relation  for both  sets  of  measurements,  using
\mstar\ fixed to  the model we find best fit  parameters for the slope
to be \phiredevoslopemfix\ for the  red population, consistent with no
evolution.  On the other hand, the total population with a slope equal
to   \phievoslopemfix\  hints   to   a  possible   evolution  at   the
\phiallsignificancemfix$\sigma$  level, with  clusters having  a lower
density of $m^*$ galaxies at higher redshift.

As already  mentioned in \S~\ref{sec:stackedlf}, our  LF normalization
is consistent with  values in the low redshift  regime when accounting
for the high  masses of our clusters.  At high  redshift this is among
the  first  study   of  its  kind.   Our  approach   to  studying  the
characteristic galaxy  density $\phi^*$ requires good  mass estimates,
and until recently these were not available at redshifts $z\sim 1$.

\subsubsection{Evolution of the Faint End Slope $\alpha$}
\label{sec:lfalpha}
The  redshift  evolution of  the  faint  end  slope $\alpha$  for  all
galaxies   and  for   red   galaxies  is   shown   in  panels~(b)   of
Fig.~\ref{fig:evolution} and Fig.~\ref{fig:evolution_red}.   It can be
seen that $\alpha$ changes to  less negative values at higher redshift
with  \alphaallsignificance$\sigma$ and  \alpharedsignificance$\sigma$
significance, for all and red  population, respectively.  That is, for
the red  population there  is weak evidence  for fewer  low luminosity
cluster galaxies relative to high  luminosity cluster galaxies at high
redshift than in the local Universe.  The best fit linear relation has
intercept       \alphaevointer/\alpharedevointer\      and       slope
\alphaevoslope/\alpharedevoslope\   for   all  and   red   population,
respectively.

Comparing the  results from  the total  population with  low-$z$ Abell
Clusters, in bands redward of  the 4000\AA~break, we find a consistent
picture.  For example, \citet{gaidos97}  observed 20 Abell Clusters in
the $R-$band obtaining $\alpha  = -1.09 \pm 0.08$.  \citet{paolillo01}
constructed  the  LF using  39  Abell  Clusters  and found  $\alpha  =
-1.11^{+0.09}_{-0,07}$, in Gunn $r-$band.  \citet{barkhouse07} studied
57 Abell Clusters, in $R_C$ band, constructing the red, blue and total
LF.  For the  total LF they find an agreement  with $\alpha=-1$ in the
region  just  fainter  than  $m^*$  and  a  steeper  $\alpha$  as  the
photometry  gets deeper,  in the  range that  is not  covered by  this
study.  Also, \citet{piranomonte01} examined 80 Abell Clusters finding
$\alpha = -1.01^{+0.09}_{-0,07}$ in Gunn $r-$band.

At   higher   redshifts,   in    agreement   with   low-$z$   studies,
\citet{popessoII05}  used X--ray  selected samples  at redshift  $\leq
0.25$  and found  a faint  end  slope $\alpha  = -1.05  \pm 0.13$,  in
$r-$band,  for the  brighter  part of  the LF  and  with a  background
subtraction  method  similar  to  our approach.   Also,  in  the  same
redshift range, \citet{hansen05} showed qualitatively that $\alpha=-1$
is a good fit to X--ray selected clusters in $r-$band using SDSS data.

At ever  higher redshift, the  observational efforts to obtain  the LF
are  more common  in the  infrared,  as it  is expected  to track  the
stellar  mass  without great  sensitivity  to  recent star  formation.
\citet{lin06} used 27 clusters at redshifts $0  < z < 0.9$ to find the
low-redshift faint-end slope of $\alpha=-0.9$ qualitatively consistent
with their  high redshift  sample.  \citet{muzzin07a} found  a similar
slope $\alpha=-0.84\pm0.08$ with a sample  of 15 clusters at redshifts
$0.2  <  z  <  0.5$. Using  \spitzer,  \citet{mancone12b}  found  also
shallower  slopes,   with  $\alpha_{3.6\mu  m}=-0.97  \pm   0.28$  and
$\alpha_{4.5\mu\rm m}=-0.91 \pm 0.28$ in lower mass clusters or groups
at  $<z>\sim  1.35$.  Recently, \citet{chiu16b}  also  used  $Spitzer$
3.6$\mu$m to  construct the LF of  46 low mass systems,  within a wide
redshift range.  They  found an LF faint slope of  $\alpha \sim -0.9$,
within $0.1  < z  < 1.02$,  consistent with  no evolution  albeit with
large error bars.

The literature points  to little evolution of  $\alpha$, with high-$z$
cluster LFs  being shallower  (albeit with  redder rest  frame bands).
Our results show $\alpha$ evolution for the full population consistent
with no  evolution up to redshift  1.1. For the red  population, there
are several studies  that show that the red sequence  LF slope evolves
strongly    with    shallower    $\alpha$    at    higher    redshifts
\citep[e.g.,][]{delucia04,goto05,tanaka05,barkhouse07,stott07,gilbank08,rudnick09}. Our
findings show an evolutionary trend  on $\alpha_{\rm red}$ as reported
in   previous  works   at  the   \alpharedsignificance$\sigma$  level.
Nevertheless,    a    closer    inspection    of    panel    (b)    of
Fig.~\ref{fig:evolution_red} seems  to show that at  the high redshift
end  the trend  is dominated  by a  single cluster,  SPT-CLJ0102-4915,
observed with VLT  and with a background subtraction  done with COSMOS
data.  To  estimate the impact of  the cluster we perform  a bootstrap
resampling of the data, revealing a similar positive trend in $\alpha$
evolution of \alphaslopebootstrap\  $\pm$ \alphaslopeerrbootstrap, but
with a  lower significance (\alphaslopesignificancebootstrap$\sigma$).
A larger sample  of SZE selected clusters is needed  to strengthen our
results, especially in the high redshift end.

\begin{figure*}
\begin{center}
\includegraphics[width=1\textwidth]{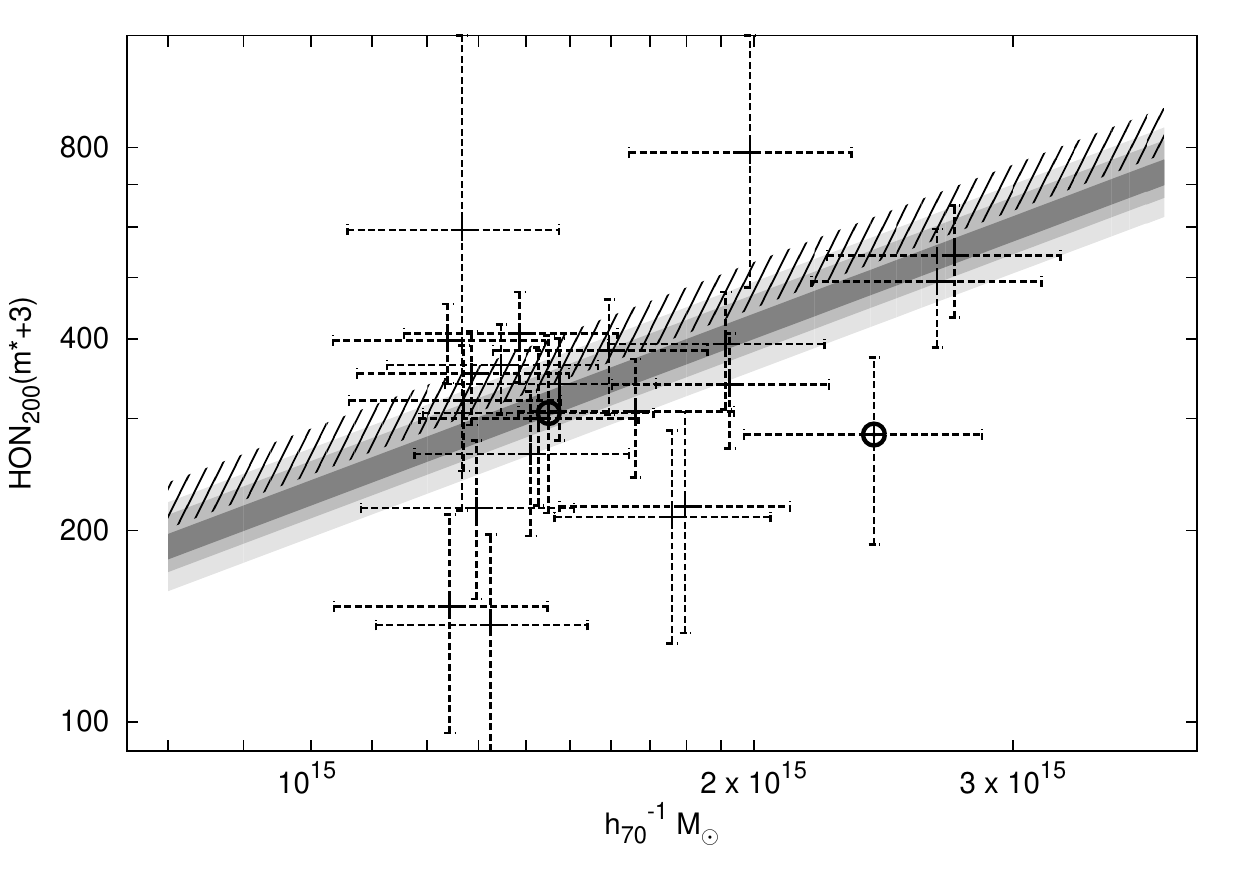}
\caption[Halo Occupation  Number for  a fixed  slope]{We plot  the HON
  constructed  using the  LF extracted  from the  band redward  of the
  4000\AA~break  versus cluster  mass,  as extracted  from the  SPT-SZ
  survey.   The VLT  data is  highlighted with  circles. Shaded  areas
  correspond to  the 1, 2 and  3 $\sigma$ errors in  the normalization
  given a fixed slope of $\gamma  = 0.87$.  We find a normalization at
  $M_{\rm pivot}=10^{15}$M$_\odot$  of \HONNORMPIVOT~(1$\sigma$) lower
  than with $267 \pm 22$ from  L04 (1 $\sigma$ error contours shown as
  diagonal lines).\label{fig:HON} }
\end{center}
\end{figure*}

\subsection{Halo Occupation Number}

We use a homogeneously selected cluster sample to characterize the HON
as  a function  of  mass and  redshift and  then  to examine  possible
evolutionary  trends.  The  Halo  Occupation  Number  is  obtained  by
integrating the Schechter Function.
\begin{displaymath}
N= 1 +  N^s, \ {\rm with}\ N^s =  V \phi^*\int^{\infty}_{y_{\rm low}} y^{\alpha}
e^{-y}\ dy
\end{displaymath}
where the first  term accounts for the  BCG, which is not  part of the
LF, $V$  is the cluster  virial volume, $y_{\rm  low}=L_{\rm low}/L_*$
and  $\alpha$  and  $\phi^*$  are  the  values  obtained  in  previous
sections.  To compare to previous studies such as L04 we integrate the
LF to $m^*+3$ .

As can be seen in Fig.~\ref{fig:HON} the range of masses in our sample
is  quite  small,  and  so  it  is  not  possible  to  constrain  both
normalization and slope of the HON-mass relation.  Therefore, we adopt
the slope of 0.87 reported in the literature for a large sample of low
redshift clusters (L04).  With this slope, we  measure a normalization
of \HONNORMPIVOT\  (1$\sigma$ uncertainties), which is  lower than the
value found by L04 of $267 \pm 22$.

Furthermore, we  look for  possible evolution  by examining  the ratio
between our  measured HON and the  value at the same  mass obtained at
low redshift (L04).  In this analysis  we enhance the HON errors using
the mass uncertainties and the adopted  mass slope of 0.87.  Fitting a
linear    relation    in    log    space    (see    panel    (d)    in
Fig.~\ref{fig:evolution})     we      obtain     \HONevoslope\     and
\HONevointer\  for the  slope and  intercept, respectively.   Thus, we
find evidence at the \HONallEvolSignificance$\sigma$ level that galaxy
clusters at high redshift have fewer galaxies per unit mass to $m^*+3$
than  their  low-z  counterparts.   This  result  is  consistent  with
\citet{capozzi12},  where  the  HON  was   found  to  exhibit  a  mild
evolution.

One concern we have is that our VLT cluster LFs suffer from background
over-subtraction.  As we mentioned in \S~\ref{sec:sims} we use the NFW
profile  to correct  for cluster  galaxies in  the defined  background
region.   While in  the  non-VLT  data the  background  is defined  at
$r>1.5r_{200}$, for  the VLT  clusters it  is defined  at $r>r_{200}$,
which means  that a larger  correction is  being made to  the measured
background.   This correction  is  at  the $12\pm  4$\%  level for  23
clusters, while for SPT-CLJ2106-5844 at  z = 1.131, this correction is
at the $49\%$ level.  In the case of SPT-CLJ0102-4915 at $z = 0.87$ an
external  background is  used  (COSMOS), rendering  a  much lower  HON
compared   to   the   best   fit   (see   circled   right   point   in
Fig.~\ref{fig:HON}), although not constituting  a clear outlier.  This
suggests  that  the contamination  corrections  we  apply to  the  VLT
backgrounds are  not resulting in  biased HON estimates.   However, in
the complementary analysis of \citet{hennig16}, which uses DES imaging
data over  large regions  so that the  background subtraction  is less
problematic, there  is a statistically lower  significant evidence for
redshift evolution.

\section{Conclusions}
\label{sec:conclusions}

We have  studied a cluster  sample consisting  of the 26  most massive
galaxy  clusters selected  in  the 2500  deg$^2$  SPT-SZ survey.   The
masses range  between M$_{200,c} =  1.2 \times 10^{15}  $M$_\odot$ and
$2.7 \times 10^{15} $M$_\odot$, and  the redshift range is broad $0.10
\lesssim  z \lesssim  1.13$.  We  use the  SZE based  cluster mass  to
define the virial region within  which we study the optical properties
such  as the  radial profile,  the  luminosity function  and the  Halo
Occupation Number.

The stacked radial profile of the whole sample is well described by an
NFW model with a concentration of \RPallstacksol~which is low compared
to the majority  of the results found in  the literature.  Differences
between  our study  and previous  works  include the  mass range,  the
redshift extent  and the selection.   Using SDSS clusters  and groups,
\citet{hansen05} found  a strong inverse correlation  between mass and
concentration which may explain the  lower concentration we see in our
high  mass  sample, although  \citet{budzynski12}  did  not find  such
correlation using a  different radius definition on  the same dataset.
Furthermore, our  low concentration measurement is  driven by clusters
in  the  higher  redshift  bin,  which are  not  represented  in  most
previously published samples \citep{carlberg97,lin04a,budzynski12}.  A
more similar sample  to compare to our higher redshift  sample is that
in \citet{capozzi12}.  Although  having a lower average  mass than our
sample,  the  concentration  found is  $c_{\rm  g}=2.8^{+1.0}_{-0.8}$,
which is consistent with our findings.

We also stack  the red galaxy population-- defined using  a colour bin
of $\pm0.22$  centered on the  red sequence at each  redshift, finding
them  to be  more concentrated  than the  total population  at $c_{\rm
  g,red}=$\RPredstacksol.   A  higher  NFW concentration  in  the  red
population is  expected from the  observed radial distribution  of the
fraction of  red galaxies,  which increases toward  the center  of the
cluster \citep[e.g.,][]{goto04,verdugo12,ribeiro13,gruen13a}.

Evidence for the redshift evolution  of the concentration for the full
population is weak at the \CGallevosignificance$\sigma$ level.  In the
case of  the red sequence  population the redshift evolution  index is
\CGredevosolslope,  which  provides  evidence  for  evolution  at  the
\CGredevosignificance $\sigma$  level, a  trend qualitatively  in line
with DM only  simulations \citep[e.g.,][]{duffy08}. As can  be seen in
Fig.~\ref{fig:cg_evol},  this  result  is strongly  dependent  on  the
lowest redshift cluster  bin.  A larger sample, in  number of clusters
and area  coverage, is  required to further  examine this  issue.  The
Dark Energy Survey (DES) is ideally suited to address this question.

The stacked total luminosity function (LF)  is well fit by a Schechter
function with Schechter parameters: \alphaallsol\ and \phiallsol.  The
faint end  slope is found  to be  consistent with previous  studies of
local                                                         clusters
\citep[e.g.,][]{gaidos97,paolillo01,piranomonte01,barkhouse07}     and
cluster          at         somewhat          higher         redshifts
\citep[e.g.,][]{popessoII05,hansen05}.  Also, the $\phi^*$ value found
is  somewhat  lower  than   previous  work  (L04,  $\phi^*=4.00\pm0.16
~h^3_{70}~ \rm{Mpc}^{-3}$ for the case of the 25 most massive systems,
which has  a median mass  lower than ours), although  when considering
the $\phi^*$-$\alpha$  covariance they  are in  qualitative agreement.
The stacked  red-sequence LF  (rLF) is  also well  fit by  a Schechter
function with Schechter parameters: \alpharedsol\ and \phiredsol.  The
$\alpha_{\rm   rs}$  found   is  consistent   with  previous   studies
\citep{gilbank08,rudnick09}.

We also fit the LF of individual clusters using \alphaallsol\  from the stacked 
result to study the single cluster  \mstar\ evolution. We use the band
which  probes  the portion  of  the  galaxy  spectrum redward  of  the
4000~\AA\ break over  the full redshift range.   The \mstar\ behaviour
with redshift yields a slope  of \mstarevoslope, indicating  that the
evolution of the characteristic luminosity in this uniformly selected
sample  does not deviate from  the CSP model to which we compare.
This model is an exponential burst at $z=3$ with decay time of 0.4~Gyr
and a Salpeter IMF.  This is  broadly in agreement with previous work,
which has  shown cluster  galaxies  are generally well modeled  by  a
passively evolving stellar population  that formed at redshift $z>1.5$
\citep[e.g.,][]{depropris99,lin06,andreon08,mancone10}.

We used  this result, fixing \mstar\  to the CSP model  predictions in
the LF  fit to explore  the $\alpha$  and $\phi^*$ evolution.   In the
case  of  $\alpha$  evolution,  we find  a  slope  of  \alphaevoslope,
indicating no  evolution.  In the  rLF $\alpha_{\rm red}$ case,  it is
found to evolve  as \alpharedevoslope, a \alpharedsignificance$\sigma$
level evidence for  low redshift clusters having a  steeper faint end,
indicating  an  evolution  in  the  ratio  of  bright/dwarf  galaxies.
Nevertheless,  this  significance  is  greatly  reduced  if  we  do  a
bootstrap           resampling          of           the          data
(\alphaslopesignificancebootstrap$\sigma$).      The     normalization
$\phi^*/E^2(z)$  measurements  provide   no  significant  evidence  of
redshift evolution  when \mstar\  is fixed  to the  model for  the red
population,  and some  evidence (\phiallsignificancemfix$\sigma$)  for
evolution of the total population.

We measure  the HON, the number  of galaxies within the  virial region
more luminous  than $m^*+3$,  comparing it to  the literature  using a
$N\propto M^\gamma$ parametrization, and  probing for redshift trends.
Due to the small mass range in  our sample, a simultaneous fit of both
the normalization and  the slope does not  provide useful constraints.
Therefore, we adopt a slope of $\gamma=0.87$ from the literature (L04)
and  fit   for  the  normalization.    We  find  a   normalization  of
\HONNORMPIVOT\ at  a mass  $M_{200}=10^{15}$M$_\odot$, which  is lower
than  the normalization  of  $267 \pm  22$, found  in  L04 from  local
clusters.

HON evolution with redshift is found  to have a slope of \HONevoslope,
providing   some  evidence   (\HONallEvolSignificance$\sigma$)  of   a
preference for high redshift clusters  to be less populated than their
lower   redshift  counterparts   as  suggested   by  \citet{capozzi12}
findings.  A bigger sample is needed to investigate further the HON.

These results  are to be further  tested as the Dark  Energy Survey is
completed, enabling us  to probe the galaxy  population variations not
only with redshift but also with mass.

\section*{Acknowledgements}
We acknowledge the  support by the DFG Cluster  of Excellence ``Origin
and Structure  of the  Universe'', the  Transregio program  TR33 ``The
Dark  Universe'' and  the  Ludwig-Maximilians-Universit\"at. The  data
processing has  been carried  out on the  computing facilities  of the
Computational Center for Particle and Astrophysics (C2PAP), located at
the Leibniz Supercomputer  Center (LRZ).  The South  Pole Telescope is
supported   by  the   National   Science   Foundation  through   grant
PLR-1248097.  Partial support  is  also provided  by  the NSF  Physics
Frontier  Center   grant  PHY-1125897   to  the  Kavli   Institute  of
Cosmological  Physics   at  the  University  of   Chicago,  the  Kavli
Foundation and the  Gordon and Betty Moore Foundation  grant GBMF 947.
CR  acknowledges  support  from   the  Australian  Research  Council's
Discovery  Projects scheme  (DP150103208).  This  paper includes  data
gathered  with the  Blanco  4-meter telescope,  located  at the  Cerro
Tololo  Inter-American Observatory  in  Chile, which  is  part of  the
U.S. National Optical Astronomy Observatory,  which is operated by the
Association of  Universities for  Research in Astronomy  (AURA), under
contract with the  National Science Foundation.  Other  data come from
the European Southern Observatory telescopes  on La Silla and Paranal.
We are very grateful for the efforts of the CTIO, La Silla and Paranal
support staff without whom this paper would not be possible.

\bibliographystyle{mnras}
\bibliography{W11_gal_pop_mnras_format}
\end{document}